\documentclass[12pt]{iopart}
\usepackage{graphicx}

\def\gtwid{\mathrel{\raise.3ex\hbox{$>$\kern-.75em\lower1ex\hbox{$\sim$}}}}
\def\alt{\mathrel{\raise.3ex\hbox{$<$\kern-.75em\lower1ex\hbox{$\sim$}}}}
\def\agt{\mathrel{\raise.3ex\hbox{$>$\kern-.75em\lower1ex\hbox{$\sim$}}}}

\newcommand{\be}{\begin{equation}}
\newcommand{\ee}{\end{equation}}

\newcommand\Rey{\mbox{\textrm{Re}}}  

\newcommand\Pra{\mbox{\textrm{Pr}}}  
\newcommand\Ra{\mbox{\textrm{Ra}}}  
\newcommand\Nu{\mbox{\textrm{Nu}}}  

\begin{document}

\title[Heat transport by turbulent convection for $\Ra\ \alt 10^{15}$]{Heat transport by turbulent Rayleigh-B\'enard convection for $\Pra\ \simeq 0.8$ and $3\times 10^{12} \alt \Ra\ \alt 10^{15}$: Aspect ratio $\Gamma = 0.50$}

\author{Guenter Ahlers$^{1,6}$, Xiaozhou He$^{2,6}$, Denis Funfschilling$^{3,6}$ and Eberhard Bodenschatz$^{2,4,5,6}$}

\address{$^{1}$Department of Physics, University of California, Santa Barbara, CA 93106, USA}
\address{$^{2}$Max Planck Institute for Dynamics and Self-Organization (MPIDS), 37077 G\"ottingen, Germany}
\address{$^{3}$LSGC CNRS - GROUPE ENSIC, BP 451, 54001 Nancy Cedex, France}
\address{$^{4}$Institute for Nonlinear Dynamics, University of G\"ottingen, 37077 G\"ottingen, Germany}
\address{$^{5}$Laboratory of Atomic and Solid-State Physics and Sibley School of Mechanical and Aerospace Engineering, Cornell University, Ithaca, NY 14853, USA}
\address{$^{6}$International Collaboration for Turbulence Research}

\begin{abstract}
We report experimental results for heat-transport measurements, in the form of the Nusselt number \Nu, by turbulent Rayleigh-B\'enard convection in a cylindrical sample of aspect ratio $\Gamma \equiv D/L = 0.50$ ($D = 1.12$ m is the diameter and $L = 2.24$ m the height). The measurements were made using sulfur hexafluoride at pressures up to 19 bars as the fluid. They are for the Rayleigh-number range $3\times 10^{12} \alt \Ra \alt 10^{15}$ and for Prandtl numbers \Pra\ between 0.79 and 0.86. For $\Ra < \Ra^*_1 \simeq 1.4\times 10^{13}$ we find $\Nu = N_0 \Ra^{\gamma_{eff}}$ with $\gamma_{eff} = 0.312 \pm 0.002$, consistent with classical turbulent Rayleigh-B\'enard convection in a system with laminar boundary layers below the top and above the bottom plate. For $\Ra^*_1 < \Ra < \Ra^*_2$ (with $\Ra^*_2 \simeq 5\times 10^{14}$) $\gamma_{eff}$ gradually increases up to $0.37\pm 0.01$. We argue that above  $\Ra^*_2$ the system is in the ultimate state of convection where the boundary layers, both thermal and kinetic, are also turbulent. 

Several previous measurements for $\Gamma = 0.50$ are re-examined and compared with the present results. Some of them show a transition to a state with $\gamma_{eff}$ in the range from 0.37 to 0.40, albeit at values of \Ra\ in the range from $9\times 10^{10}$ to $7\times 10^{11}$ which is much lower than the present $\Ra^*_1$ or $\Ra^*_2$. The nature of the transition found by them is relatively sharp and does not reveal the wide transition range observed in the present work. 

In addition to the results for the genuine Rayleigh-B\'enard system, we present measurements for a sample which was not completely sealed; the small openings permitted external currents, imposed by density differences and gravity,  to pass through the sample. That system showed a sudden decrease of $\gamma_{eff}$ from 0.308 for $\Ra < \Ra_t \simeq 4\times 10^{13}$ to 0.25 for larger \Ra.

A number of possible experimental effects is examined in a sequence of Appendices; none of these effects are found to have a significant influence on the measurements.

\end{abstract}

\maketitle

\section{Introduction}

In this paper we consider turbulent convection in a fluid contained between horizontal parallel plates and heated from below (Rayleigh-B\'enard convection or RBC; for a reviews written for broad audiences see Refs.~\cite{Ka01,Ah09}; for more specialized reviews see Refs.~\cite{AGL09,LX10}). The primary purpose of the work on which we report was to search for the transition to the ``ultimate" state of turbulent convection first predicted by Robert Kraichnan \cite{Kr62} and Ed Spiegel \cite{Sp71} half a century ago. 
 
We focus on the particular case of a cylindrical sample of aspect ratio $\Gamma \equiv D/L = 0.50$ ($D = 1.12$ m and $L=2.24$ m are the diameter and height respectively) because this particular geometry was used in previous searches for this state \cite{CGHKLTWZZ89,CCCCCH96,CCCHCC97,NSSD00,NSSD00e,CCCCH01,RCCH01,RGCH05,GR08,GSBGPTR09,RGKS10} and thus enables a more direct comparison with earlier measurements. Experiments searching for the ultimate state using other values of $\Gamma$ are of course important as well and some have been carried out \cite{RGKS10,FG02,NS03,NS06a,UMS11,HFBA12}; but they are beyond the scope of this paper. The work reported here consists of measurements of the heat transport by the turbulent system. Other aspects will be discussed separately. 

We present results   that were obtained in the High-Pressure Convection Facility (the HPCF, a cylindrical sample of 1.12 m diameter) at the Max Planck Institute for Dynamics and Self-organization in G\"ottingen, Germany using sulfur hexafluoride (SF$_6$) at pressures up to 19 bars as the fluid. Early results from this work were presented in Refs.~\cite{AFB09, AFB11a,AFB11b}. A description of the apparatus was given in Ref.~\cite{AFB09}. The present paper presents new results obtained after various sample-chamber modifications to be described in this paper and is a comprehensive report on this work. A brief report of these recent results was provided in Ref.~\cite{HFNBA12}.

The HPCF is located inside a pressure vessel known as the Uboot of G\"ottingen which gets filled with the gas of choice. Thus there must be some way for the gas to enter or leave the HPCF. Originally a small gap of average width about one mm was permitted for this purpose to remain between the top and bottom plate and the side wall of the sample \cite{AFB09}. Erroneously it was assumed that this gap was negligible because it is small compared to the 1120 mm sample diameter. This sample will be called the ``open" sample. It turned out that the results depended on the temperature difference between the fluid in the sample at a temperature $T_m$ and the fluid in the remainder of the Uboot at $T_U$ \cite{AFB11}. For that reason the HPCF was modified by sealing the bottom plate to the side wall, but leaving the gap between the top plate and the side wall open. This system will be called the ``half-open" sample. There still was a major difference  between the results for $T_m > T_U$ and $T_m < T_U$, although these results differed from those of the closed sample.  Thus, as a final measure, both  the top and the bottom plates were sealed to the side wall, and a 25 mm diameter tube was installed to permit the gas to enter the HPCF. One end of the tube was flush with the inside of the side wall, and the other terminated in a remotely controlled valve. The sample could thus be filled while the valve was open, and then during measurements the valve could be closed. This sample will be called the ``closed" sample. Results for all three versions will be presented in this paper.

In the next section we shall define the parameters needed to describe this system (Sec.~\ref{sec:sys_paras}). We shall then, in Sec.~\ref{sec:RBC}, outline the main features of turbulent convection as they are now understood. First, in Sec.~\ref{sec:classical}, we describe the classical state of turbulent convection which exists below the transition to the  ultimate state with turbulent BLs.   This will be followed in Sec.~\ref{sec:ultimate} by a description of what is known or expected for the ultimate state.
This introductory material will be followed in Sec.~\ref{sec:apparatus} by a brief discussion of  the apparatus modifications used in this work. A detailed description of the main features was presented before \cite{AFB09}. Section~\ref{sec:Results} presents a comprehensive discussion of our results and of the results of others at large \Ra\ for cylindrical samples with $\Gamma = 0.50$. It is followed by a Summary of our work in Sec.~\ref{sec:summary}.

In a sequence of six Appendices we discuss a number of experimental issues which might be of lesser interest to the general reader but which are of considerable importance to the specialist. First, in \ref{app:tilt} we compare measurements for three different angles of tilt of the sample axis relative to gravity. The effect of a tilt was studied before by several groups, with varying results \cite{CCL96,CCS97,CRCC04,SXX05,ABN06,RGKS10}. A tilt is used at times  by experimentalists to give the large-scale circulation in the sample a preferred azimuthal orientation. This was our motivation as well. We show that it had no discernible effect on \Nu. 
In \ref{app:NOB} we demonstrate that non-Oberbeck-Boussinesq effects \cite{Ob79,Bo03} have only a very minor effect on \Nu\ which can be seen only at the largest values of $\Delta T$ near 20 K. 
In \ref{app:NS_NOB} we present values of the parameter $\xi$ introduced recently by Niemela and Sreenivasan \cite{NS10} to describe a special non-Oberbeck-Boussinesq effect which apparently occurs near critical points. We find that for our work $\xi \simeq 1.3$, indicating that the effect discussed in Ref.~\cite{NS10} (which occurs for small or negative $\xi$) is not expected to be relevant to the present work.
In \ref{app:plate_T} we give results for the horizontal temperature variations in the top and bottom plate of the sample and demonstrate that they do not influence the measured values of \Nu. 
In \ref{app:closed} we show that the ``closed" sample really was completely sealed.
In \ref{app:sideshield} we present data for the influence of a mismatch between the temperature of the sample side shield and the sample mean temperature $T_m$, and show that the side-shield temperature-regulation is good enough for these effects to be negligible.  
Finally, in \ref{app:data}, we give a complete list of our data in numerical form.

\section{The system parameters}
\label{sec:sys_paras}

For turbulent RBC in cylindrical containers there are two parameters which, in addition to $\Gamma$, are expected to determine its state. They are the dimensionless temperature difference as expressed by the Rayleigh number
\be
\Ra \equiv \frac{\alpha g \Delta T L^3}{\kappa \nu}
\label{eq:Ra}
\ee 
and the ratio of viscous to thermal dissipation as given by the Prandtl number
\be
\Pra \equiv \nu/\kappa\ .
\label{eq:Pr}
\ee 
Here $\alpha$ is the isobaric thermal expansion coefficient, $g$ the gravitational acceleration, $\kappa$ the thermal diffusivity, $\nu$ the kinematic viscosity, and $\Delta T \equiv T_b - T_t$ the applied temperature difference between the bottom ($T_b$) and the top ($T_t$) plate. 

In the present paper we present measurements of the heat transport. These results are presented in the form of the scaled effective thermal conductivity known as the Nusselt number which is given by
\be
\Nu \equiv \frac{Q L}{A \Delta T \lambda}\ .
\label{eq:Nu}
\ee
Here $Q$ is the applied heat current, $A = D^2\pi/4$ the sample cross-sectional area, and $\lambda$ the thermal conductivity. The measurements cover the range  $10^{12} \alt \Ra \alt 10^{15}$ and are for \Pra\ ranging from 0.79 at the lowest to 0.86 at the highest \Ra. 

All fluid properties needed to calculate \Ra, \Pra, and \Nu\ were evaluated at the mean temperature $T_m = (T_t+T_b)/2$ of the sample. They were obtained from numerous papers in the literature, as discussed in Ref.~\cite{LA97}.

\section{The characteristics of turbulent RBC}
\label{sec:RBC}
 
\subsection{The classical state.}
\label{sec:classical}
 
A ``classical" state of RBC  exists below a transition {\it range} to an ``ultimate" state; the transition range extends over more than a decade, approximately from $\Ra^*_{1}$ to $\Ra^*_{2}$ \cite{HFNBA12}. For simplicity of discussion  we shall characterize this range by $\Ra^*$, taken to lie perhaps somewhere near the middle of the range,   which, for the parameters of our work, is about  $10^{14}$ \cite{GL02,HFNBA12}. For $\Ra \alt \Ra^*$ the heat transport in this system is controlled by laminar thermal boundary layers (BLs), one just below the top and the other just  above the bottom plate. The value of $\Ra^*$ has been the subject of discussion for some time, and a major issue at the forefront of the field is the nature of the state above $\Ra^*$. Estimates of $\Ra^*$ are not very accurate; a reasonable argument  \cite{GL02} yielded $\Ra^* \simeq 10^{14}$ or so  for $\Pra \simeq 1$, although another estimate \cite{NS03} gave a value closer to $10^{12}$.  

For $\Ra < \Ra^*$  nearly half of $\Delta T$ is found across each BL, and the sample interior (known as the ``bulk")  has a highly fluctuating temperature which is nearly uniform in the time average \cite{BTL94,LX98}. At a more detailed level it was recognized long ago that the bulk actually sustains small temperature gradients, but the total temperature drop across it  is much smaller than that across the BLs (see, for instance, \cite{TBL93,BA07_EPL,WA11a}). Very recently it was found that these small temperature variations in the bulk take the form of a logarithmic dependence on the distance from the plates \cite{ABGHLSV12}; but the precise origin of this logarithmic variation is not yet known.

For the classical state it is well established both experimentally \cite{XBA00,AX01,XLZ02,FG02,FBNA05,NBFA05,SRSX05} and theoretically \cite{GL00,GL01,GL02,GL04} that the Nusselt number can be represented by a power law 
\be
\Nu = N_0 \Ra^{\gamma_{eff}}
\label{eq:powerlaw}
\ee
with the effective exponent varying from about 0.28 near $\Ra = 10^8$ to about 0.32 near $\Ra = 10^ {11}$, at least when \Pra\ is close to one or larger. 

It is also well established that, in cylinders with $\Gamma \simeq 0.5$ containing a fluid with $\Pra \simeq 0.7$ and for $\Ra \alt 10^{11}$, there is a large-scale circulation (LSC) in the sample interior that takes the form of a single convection roll, with up-flow along the side wall at an azimuthal position $\theta_0$ and down-flow also along the wall but at an azimuthal position close to $\theta_0+\pi$ \cite{ABFH09,WA11a}. The LSC is bombarded by the small-scale fluctuations of the system, and may be regarded as a stochastically driven system that fluctuates intensely \cite{BA07a,BA08a}. Even at modest \Ra\ below, say, $10^{11}$ and for \Pra\ near 0.7, the LSC existence is intermittent. It frequently collapses, only to re-form again at a somewhat later time.
Whether the LSC survives at all up to $\Ra^*$ had not been clear heretofore; it was recognized (see, for instance, \cite{NS03}) that the LSC becomes less well defined as  \Ra\ increases, but concrete quantitative experimental evidence for its existence or demise  has only become available during our present work. We found that, even for $\Ra = 10^{15}$, there is evidence of its existence, but its average lifetime is short and it may be regarded more appropriately as just one of the continuum of modes contributing to the fluctuation spectrum of the system.  This will be reported in detail in a subsequent paper. 

When the fluctuations are not too vigorous, the LSC due to its horizontal flow with speed $U$ just adjacent to the top and bottom plate will establish {\it viscous} BLs adjacent to the plates. The viscous BLs may be imbedded in the thermal ones or {\it vice versa}, depending on \Pra. These BLs are laminar, albeit fluctuating \cite{ZX10,SZGVXL12}, in the classical state.

\subsection{The ultimate state.}
\label{sec:ultimate}

About half a century ago it was predicted by Kraichnan \cite{Kr62} and Spiegel \cite{Sp71} that, in the absence of  boundary layers, the Nusselt number should be proportional to $Ra^\gamma$ with $\gamma = 1/2$. This prediction is consistent with rigorous upper bounds for $\Nu(\Ra)$ obtained by Howard \cite{Ho64} and by Doering and Constantin \cite{DC96}. Although it seems difficult to construct a physical system without boundaries, the $\gamma = 1/2$ prediction was supported by direct numerical simulations (DNS) of RBC with periodic boundary conditions (BCs) in the vertical direction and forcing in the bulk \cite{LT03,CLTT05,SCLTV12}, as well as by DNS for the Rayleigh-Taylor instability \cite{BMMV09} which is expected to reveal similar phenomena and has no boundaries. Experimentally, it is noteworthy that a {\it local} heat-flux measurement in the center of a Rayleigh-B\'enard sample in the classical state ({\it i.e.} in the state with laminar BLs) yielded an exponent of 0.5 \cite{STX08}, even though the global heat flux led to $\gamma_{eff} \simeq 0.3$. 

In the presence of boundaries, Kraichnan noted that the BLs should become turbulent when \Ra\ exceeds some characteristic value $\Ra^*$. This event was expected to be induced by the shear applied to the BLs by the large-scale circulation, or if none exists, by the vigorous turbulent fluctuations in the sample interior. It was expected to occur when the shear Reynolds  number 
\be
\Rey_s \equiv U \lambda_u / \nu
\label{eq:Re_s}
\ee
determined by $U$ and the BL thicknesses $\lambda_u$ exceeded a typical value somewhere in the range from about 200 to 400 \cite{LL87}. When there is no LSC because fluctuations dominate, one would expect the fluctuations to take on its role and generate shear, with their root-mean-square velocity $V$ near the BLs taking the role of $U$ in Eq.~\ref{eq:Re_s}. In that case the characteristic size of fluctuations will cover a range, roughly from $D$ down to smaller lengths, and will be intermittent in time. One then would expect the turbulent shear layers to be more localized laterally in space, as well as in time. In Kraichnan's  considerations he assumed that the viscous and thermal BLs would undergo the shear-induced transition at the same value of \Ra. Even in the presence of rigid top and bottom plates the prediction for the large-\Ra\ asymptotic state then was still $\Nu \sim \Ra^{1/2}$. However, Kraichnan \cite{Kr62} predicted that, due to the turbulent BLs, there would be logarithmic corrections to this power law. 

Recently Grossmann and Lohse considered the possibility that the thermal and viscous BLs may undergo the turbulent shear transition at different values of \Ra\ or simultaneously, and derived the consequences of transitions in one or the other or both \cite{GL11}. The ultimate state would then correspond to the case where both the viscous and the thermal BLs become turbulent (see Sec. III.C of \cite{GL11}; we will reserve the notation $\Ra^*$ for this case). 

In analogy to the logarithmic velocity profiles in turbulent shear flows first considered by von K\'arm\'an \cite{Ka30} and Prandtl \cite{Pr32} (for a recent review, see \cite{MMMNSS10}),  Grossmann and Lohse \cite{GL11} predicted that the turbulent BLs would extend throughout the sample, replacing the bulk by a temperature profile that varies logarithmically with the distance from the plates. For the Boussinesq system \cite{Ob79,Bo03} the two profiles, one coming from the top and the other from the bottom plate, would then meet at the horizontal mid-plane of the cell. Logarithmic temperature profiles have indeed been observed in recent measurements for the ultimate state \cite{ABGHLSV12}; but since they were found for the classical state as well, it remains unclear to what extend this finding supports the prediction. In conjunction with the viscous and thermal sublayers near the plates (which survive above $\Ra^*$ because of the boundary conditions at the solid-fluid interface) 
the extended turbulent BLs lead to logarithmic corrections to the asymptotic power law for \Nu. The Grossmann and Lohse prediction for these logarithmic corrections differs from the original form of the logarithms given by Kraichnan. 
However, for either prediction the corrections vary only slowly with \Ra, and in experimentally accessible \Ra\ ranges one expects an effective power law with an effective exponent $\gamma_{eff} \simeq 0.38$ to 0.40. The asymptotic regime where the  effective exponent has essentially reached 1/2 is well out of reach of any conceivable experiment.

An important question was the value of $\Ra^*$. Since the Reynolds number $\Rey = UL/\nu$ of the global LSC (and thus $U$ and $\Rey_s$ in Eq.~\ref{eq:Re_s}) decrease with increasing \Pra\ at constant \Ra\ (see, for instance, Ref. \cite{GL02}) it follows that $\Ra^*$ increases with increasing \Pra. The  value of $\Rey_s^*$ depends on the nature and amplitude of prevailing perturbations, but is estimated to be in the range from 200 to 400 \cite{LL87}.  For $\Pra \simeq 1$ these considerations led to $\Ra^* = {\cal O}(10^{14})$ \cite{GL02}; but other estimates \cite{NS03}  yielded lower values.

A notable recent success in the search for the ultimate state has been  achieved with turbulent Couette-Taylor (CT) flow \cite{GHBSL11,HGGSL12} (it had been shown on the basis of its equations of motion that it should undergo an ultimate-state transition that is analogous  to that of RBC \cite{EGL00}). In the CT case the shear is applied directly to the fluid by concentric rotating cylinders, and thus is much more effective in driving the BLs into the turbulent state than is the shear in RBC which is generated as a secondary effect by the buoyancy-induced LSC or the fluctuations. The CT measurements yielded an effective exponent of 0.38 for the corresponding variables, consistent with 1/2 and the predicted logarithmic corrections \cite{GL11}.

For RBC the situation is less clear. In order to reach exceptionally high \Ra, two groups used fluid helium near its critical point at temperatures of about 5 K and pressures of about 2 bars. One of them \cite{CCCHCC97,CCCCH01}, at the time located at Grenoble, reported to have found the ultimate regime, and cited a value $\Ra^* \simeq 10^{11}$. We shall refer to these results as the "Grenoble" data. A major puzzle created by these results is that one can estimate that the data imply $\Rey_s^* \simeq 100$ or less, and this seems  too low for any BL shear instability. 

In a second nominally equivalent investigation near the critical point of helium Niemela et al. \cite{NSSD00} made measurements of \Nu\ up to $\Ra \simeq 10^{17}$. They found that $\Nu \propto \Ra^{0.32}$ \cite{NS06b} up to their largest \Ra, without any evidence for a transition. This work was done at the University of Oregon, and we shall refer to these results as the "Oregon" data. In this case the absence of a transition does not necessarily contradict expectations  because \Pra\ began to increase as \Ra\ exceeded about $10^{13}$, and it is plausible that $\Ra^*(\Pra)$ never was reached or resolved in that experiment.

There has been a number of additional low-temperature experiments intended to clarify the situation; we refer to a recent review \cite{AGL09} for a detailed discussion of these measurements. For completeness we mention a comprehensive recent article by  Roche et al. \cite{RGKS10} which examines the influence of the nature of the side walls, of $\Gamma$, of \Pra, and of several other factors which seem to influence the transition to a state with $\gamma_{eff}$ significantly larger than 0.32. This survey concludes that a transition to the ultimate regime occurs in several experiments, but again these transitions occurred at unexpectedly low values of \Ra. The survey concludes that the transition occurs at smaller \Ra\  when \Pra\ is larger, which is opposite to the \Pra\ dependence of $\Ra^*$ expected for the shear instability. Since no LSC-induced shear instability is likely to have occurred, and since the ultimate-state predictions  are based on the assumption of turbulent BLs, it remains unclear to us how the states with $\gamma_{eff}$ much larger than 0.32 reported in Ref.~\cite{RGKS10} are related to the Kraichnan prediction \cite{Kr62} or to the states discussed by Grossmann and Lohse \cite{GL11}.
  
 \begin{table}[h]
\caption{Versions of the High-Pressure-Convection Facility (HPCF). The second column gives the material used  to construct the top and bottom plates. The third column indicates whether the top and/or bottom plate was sealed to the side wall.}
\vskip 0.1in
\begin{center}
\begin{tabular}{cccc}
Version  & Plates & Seals & runs \\  
\hline
HPCF-I & Aluminum & none & 080827 to 090313 \\
HPCF-IIa & Copper & none & 090505 to 090917 \\
HPCF-IIb & Copper & none & 090905 to 100125 \\
HPCF-IIc & Copper& bottom & 100202 to 100502 \\
HPCF-IId & Copper & bottom and top & 100612 to 100818 \\
HPCF-IIe & Copper & bottom and top & 100918 to 110919 \\
\end{tabular}
\end{center}
\label{tab:HPCF}
\end{table}

\begin{figure}
\centerline{\includegraphics[height=5in]{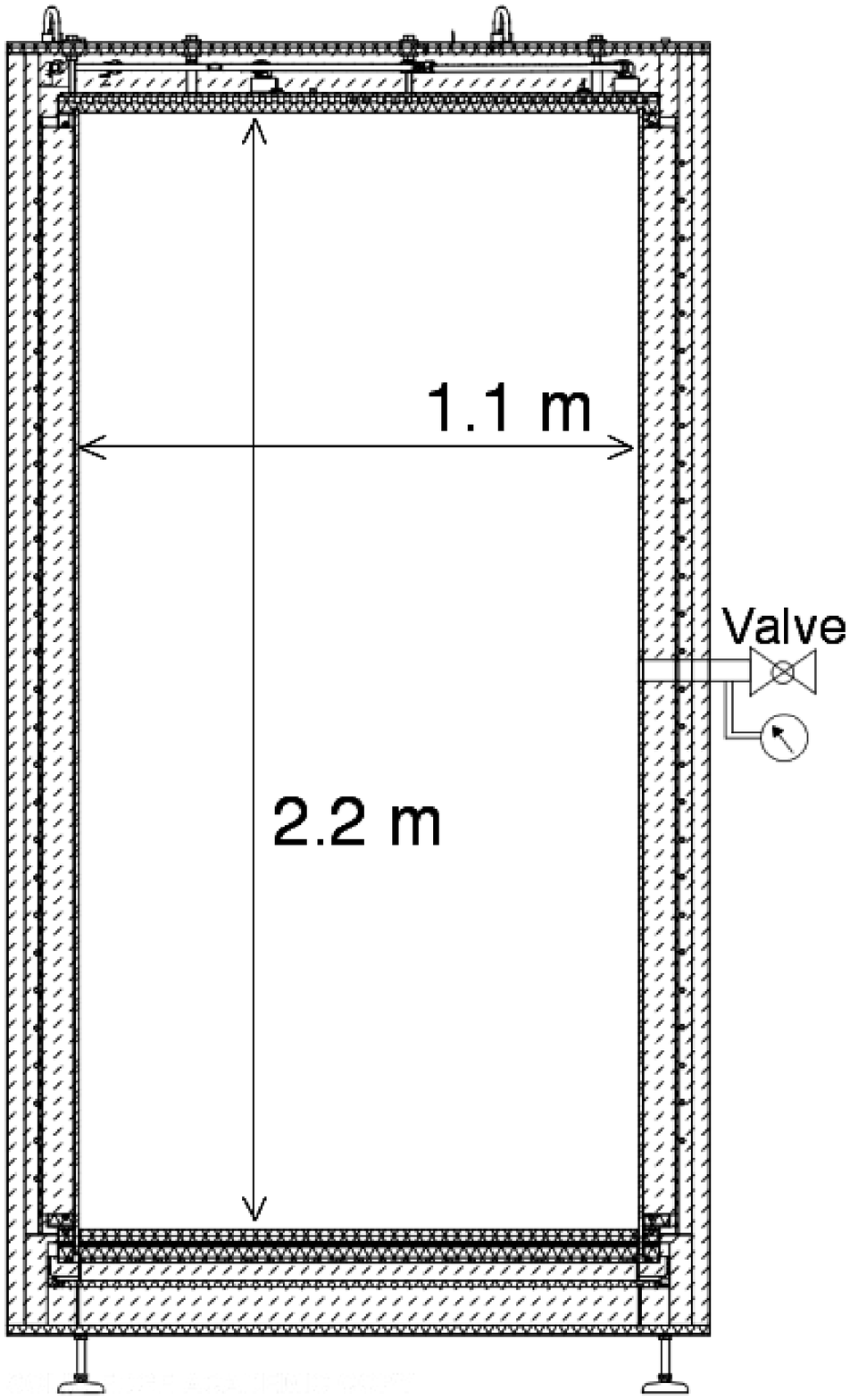}
\includegraphics[height=5in]{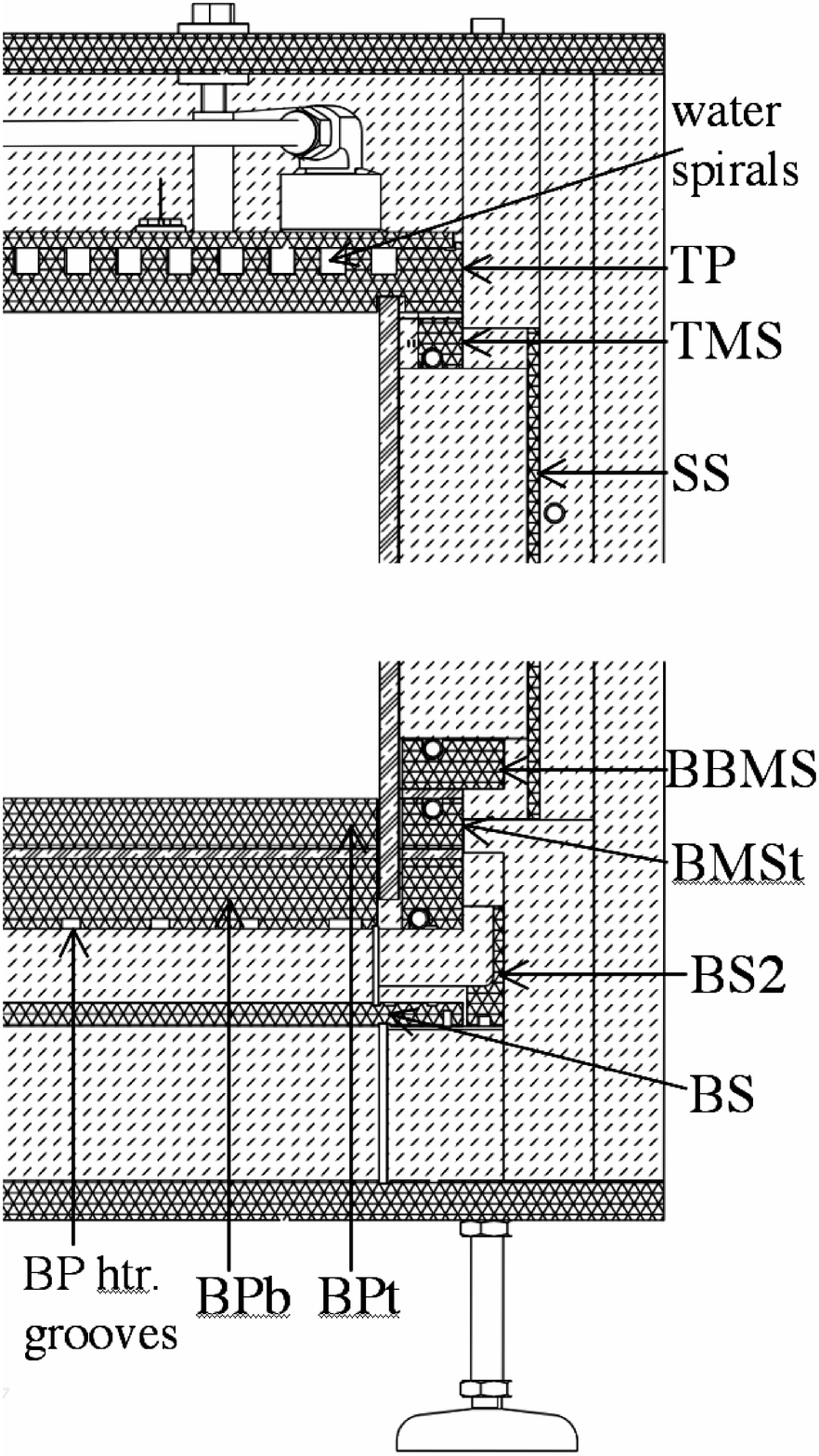}}
\caption{Left: Diagram of HPCF-IIe. Right: Detailed diagram of the top and bottom corners of HPCF-IIc to HPCF-IIe. All parts are shown to scale, except for the valve in the left part. We refer to Fig.~2 of \cite{AFB09} for a description of many features that were common with HPCF-I to HPCF-IIb. The bottom plate was a composite consisting of a bottom member ``BPb", a top member ``BPt", and a Lexan sheet between them. The bottom-plate heater was imbedded in epoxy inside the heater grooves (``BP htr grooves"). The bottom shield (``BS") was extended by adding a section (``BS2"). The bottom bulk microshield (``BBMS"), servoed at $T_m$, is new. The side shield (``SS") and top microshield (``TMS") are unchanged. The location and size of the spiral water channels (``water spirals") in the top plate (``TP") are indicated.}
\label{fig:samplecell}
\end{figure}

\section{Apparatus}
\label{sec:apparatus}

Versions HPCF-I, HPCF-IIa, and HPCF-IIb of the apparatus were described in detail in Ref.~\cite{AFB09}. A schematic diagram of these units can be found in Fig. 2 of that reference. HPCF-I had aluminum top and bottom plates, whereas HPCF-II had copper plates. HPCF-IIa and IIb differed only in the amount and type of insulation (mostly open-pore foam and aluminum-coated polyester film, see \cite{AFB09})  provided outside the sample cell, and data obtained with them showed no obvious difference. More  recently three additional modifications known as HPCF-IIc, -IId, and -IIe were developed, and corresponding schematic diagrams are shown in Fig.~\ref{fig:samplecell}. HPCF-IId and -IIe differed from HPCF-IIc only by the 2.5 cm diameter side arm and valve used to fill and empty the sealed samples, and by whether or not the side wall was sealed to the bottom and/or top plate. Table \ref{tab:HPCF} lists the major differences between the six versions, as well as the identifications of the runs\footnote{The run numbers had the structure ``yymmdd".} performed in each.

All samples had an internal height $L = 2240\pm 2$ mm and diameter  $D = 1122\pm 2$ mm. The measurements to be discussed here were made in HPCF-IIb to -IIe. 
All samples had a Plexiglas side wall of 0.95 cm thickness and several thermal shields. The entire sample was immersed in a high-pressure vessel, known as the Uboot of G\"ottingen, that could be filled with various gases, including sulfur hexa-fluoride (SF$_6$), up to a pressure of $P=19$ bars.  As shown in Fig.~\ref{fig:samplecell}, all samples had a composite bottom plate consisting of a bottom (BPb) and a top (BPt) member made of copper (aluminum for HPCF-I) and a 5 mm thick layer of Lexan sandwiched between them. The composite was glued together with very thin layers of degassed Stycast 1266 epoxy. The temperature difference across this composite, together with the composite conductance, was used to infer the heat current $Q$ that entered the sample at the bottom. The underside of the bottom member of the composite was heated electrically by a heater immersed in epoxy in grooves (``BP htr grooves"). The top plate (TP) was cooled by a water circuit consisting of two pairs of spirals.  The pairs were in parallel, and the flows in the two members of a given pair were anti-parallel. Remaining horizontal thermal gradients in the TP and the BPt are discussed in Appendix D.
 
The various shields which prevented parasitic heat losses from the sample cell were discussed in detail in \cite{AFB09}. Starting with HPCF-IIc we added two more shields. the bottom shield, which is always servoed at the temperature of the bottom member BPb of the composite,  was extended by adding the section BS2 (see Fig.~\ref{fig:samplecell}). A more significant addition was the ``Bottom Bulk Micro Shield" BBMS. It was servoed at $T_m$ and thus minimized vertical thermal gradients in the space between the side wall and the side shield (SS). Prior to the addition of BBMS there was a vertical temperature drop from the BPt temperature at BMSt to $T_m$ at TMS, which is approximately equal to $\Delta T/2$ and thus generated a Rayleigh number about equal to half the sample \Ra. Even though the space between the side wall and the SS was filled with foam and polyester film, convection was believed to have been induced in this space during runs at the larger values of \Ra.  

HPCF-IIc was identical to HPCF-IIb, except that a seal consisting of  silicone adhesive was applied to the inside corner between the side wall and the top of the bottom-plate composite along the entire periphery. It is expected that this seal will prevent any flow through the small gap, of width about 1 mm, between the side wall and the bottom plate. A similar gap between the side wall and the top plate was left open since fluid had to be allowed to enter or leave the cell as the temperature or pressure was changed.

HPCF-IId and HPCF-IIe consisted of a completely sealed system, with no gaps between the top or bottom plates and the side wall. A tube of inside diameter 2.5 cm was installed and entered the side wall at half height. Its termination was flush with the inside of the side wall, without any protrusion into the convection chamber. Outside the convection chamber this tube contained a remotely operated ball valve. A small-diameter ($\simeq$ 3 mm) tube led from the sample side of the 2.5 cm diameter tube to a location outside the Uboot where it was connected to the pressure gage. Thus, the actual sample pressure could be monitored. At each set point of the experiment the system was equilibrated with the valve open for about six hours. The valve was then closed, permitting measurements on a completely sealed system. The fill tube had two side arms with an additional valve in each. One opened  when the pressure difference $P_U - P$ between the Uboot and the sample exceeded 25 mBar; the other opened when this pressure difference was less than -25 mBar. We note that a pressure difference of 25 mBar leads to a force of about  250 kg acting on the top and bottom plates. When for instance this force exceeds the weight of the top plate, then this plate will lift up and damage will be done to the instrument. In order to keep $|P_U - P|$ sufficiently small, filling and emptying of the Uboot and sample was done very slowly, over a period of a day or two depending on the desired pressure. Measurements of the sample pressure under various conditions showed that the sample was indeed sealed, as discussed in Appendix E.

Nusselt-number measurements were based on temperatures determined with fifteen thermometers, five each in the two members BPb and BPt of the bottom-plate composite and five in the top plate TP. Each set of five consisted of one thermometer placed at the plate center and four, positioned equally spaced azimuthally, at a distance of $0.42D$ from the center. The thermometers in the BPt and the TP were located about 1 mm from the fluid-copper interface. The three sets of five thermometers were used to obtain the averages $T_{BPb}, T_{BPt},$ and $T_{TP}$. $T_{BPb}$ and $T_{BPt}$ were used to calculate the heat current entering the sample. $T_{BPt}$ and $T_{TP}$ were used to obtain $\Delta T$ and \Ra. In a typical run both $T_{BPb}$ and $T_{TP}$ were regulated at a specified setpoint.

A small correction to \Nu\ was made for the sidewall conductance \cite{Ah00,RCCHS01}.  This correction was about 1.4\% for $\Ra \simeq 10^{15}$ and about 3.5\% for $\Ra \simeq 5\times 10^{12}$. Neglecting this correction changed the exponent obtained from a power-law fit to the data for $\Nu(\Ra)$ by about -0.003. Estimates and a comparison of measurements with copper and aluminum plates \cite{FBA09} showed that corrections for the finite conductivity of the top and bottom plates \cite{Ve04,BNFA05} were negligible.

\section{Results.}
\label{sec:Results}

\subsection{Closed sample.}
\label{sec:closed}

\subsubsection{The broad overview.}

In this subsection we present results for the closed sample HPCF-IIe which is our best approximation to the idealized RB system. In Sect.~\ref{sec:compare} we compare these results with previous measurements. Then, in Sec.~\ref{sec:open}, we discuss the measurements for the open sample where the RB system may be perturbed by an additional current entering or leaving the sample through the narrow gap between the top and bottom plates and the side wall because of the chimney effect. Finally, in Sec.~\ref{sec:halfopen}, we consider the case where only the bottom of the sample is sealed (HPCF-IIc) while a gap remains between the top plate and the side wall (the ``half-open" sample).

\begin{figure}
\centerline{\includegraphics[width=4in]{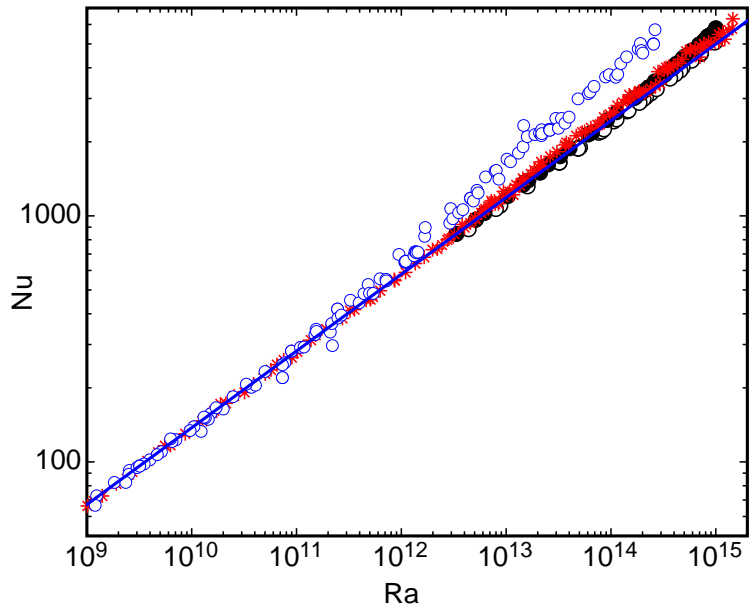}}
\caption{The Nusselt number \Nu\ as a function of the Rayleigh number \Ra\ on logarithmic scales. Solid black circles: This work, $T_m - T_U < -2$K. Open black circles: This work, $T_m - T_U >  2$K. Red stars: the Oregon data \cite{NSSD00,NSSD00e,NS06b}. Open blue circles: the Grenoble data \cite{CCCHCC97,CCCCH01}. Blue solid line: the power law $\Nu = 0.1044 \Ra^{0.312}$.}
\label{fig:Nu_of_Ra}
\end{figure}

Results of the \Nu\ measurements for the closed sample (HPCF-IIe) were reported briefly  before \cite{HFNBA12}. It was found that they depended slightly on $T_m - T_U$, but much less so than the data for the open or the half-open sample. We have been unable to determine the reason for this dependence which persisted in spite of the many thermal shields and the foam and foil insulation that were provided (see \cite{AFB09}, and  Sec.~\ref{sec:apparatus} and Fig.~\ref{fig:samplecell} above).

In Fig.~\ref{fig:Nu_of_Ra} we show $\Nu$ as a function of \Ra\ with both axes on a logarithmic scale. The open circles are for $T_m - T_U \agt 2$ K, and the solid   black circles represent the data for $T_m - T_U \alt -2$ K. Within the resolution of this figure the open and solid circles are seen to agree quite well with each other and with the Oregon data (red stars), although small differences can be noticed on close inspection. Both our data sets and the Oregon data  differ significantly from the Grenoble data (open blue circles). A more detailed comparison with those as well as with other \cite{Wu91,RGKS10}  results is given below.

\subsubsection{The classical state.}

\begin{figure}
\centerline{\includegraphics[width=5in]{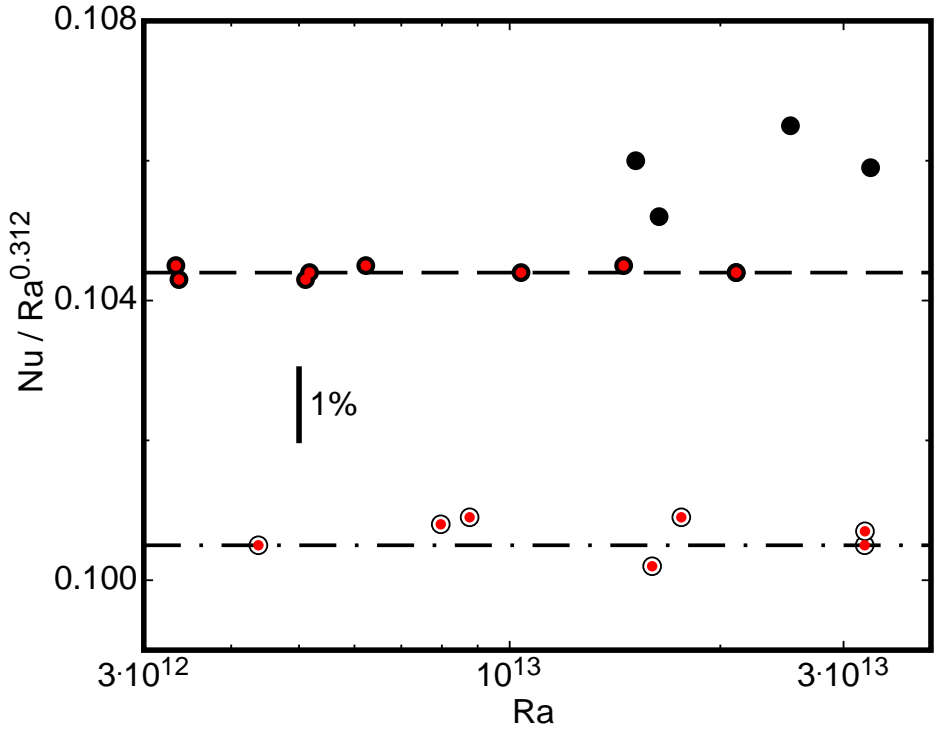}}
\caption{a): The reduced Nusselt number $\Nu/\Ra^{0.312}$ on a linear scale as a function of \Ra\ on a logarithmic scale. Solid (open) black circles: $T_m - T_U < -2$ K ($T_m - T_U > 2$ K). The points indicated by a red dot in the center were used in a least-squares fit of a power law to the data.}
\label{fig:hires}
\end{figure}

First we examine $\Nu(\Ra)$ in the classical regime in greater detail. Figure~\ref{fig:hires} is a high-resolution graph of  the data in this parameter range in the form of $\Nu/\Ra^{0.312}$ on a linear scale as a function of \Ra\ on a logarithmic scale. One sees that each of the two data sets covers a range of about a decade in the classical regime where a simple power law describes them well. A fit of the power law Eq.~\ref{eq:powerlaw} to the data points that are indicated by small red dots in their centers gave the parameter values and standard erroers (67\% confidence limits) listed in Table~\ref{tab:clasicalfitparas}. On the basis of this analysis we chose $\gamma_{eff} = 0.312$ as our best estimate of the effective exponent in the classical regime. As can be seen from the table, the uncertainty of this result due to the scatter in the data is less than 0.001. We estimate that an additional uncertainty comes from possible systematic errors of the sidewall correction, and thus the overall uncertainty of the exponent is about 0.002. This value is consistent with numerous other measurements at smaller \Ra\ and larger \Pra, and agrees quite well with the value 0.323 obtained from a numerical analysis of the Grossmann-Lohse model, Eqs.~13 and 14 of Ref.~\cite{GL01}, for $10^{12} \alt \Ra \alt 10^{13}$ and $\Pra = 0.8$ (but for $\Gamma = 1.00$).

 \begin{table}[h]
\caption{Results of least-squares fits of a power law to the data in the classical regime.}
\vskip 0.1in
\begin{center}
\begin{tabular}{ccc}
Data set  & $N_0$ & $\gamma_{eff}$ \\  
\hline
$T_m - T_U \alt -2$K & $0.1040 \pm 0.0011$ & $0.3121 \pm 0.0004$ \\
$T_m - T_U \alt -2$K & $0.1044 \pm 0.00002$   & 0.312 (fixed)			 \\
$T_m - T_U \agt ~2$K  & $0.1020 \pm 0.0037$ & $0.3116 \pm 0.0012$  \\
$T_m - T_U \agt ~2$K  & $0.1006 \pm 0.00009$  & 0.312 (fixed)			\\
\end{tabular}
\end{center}
\label{tab:clasicalfitparas}
\end{table}

\subsubsection{Transition to the ultimate state.}

\begin{figure}
\centerline{\includegraphics[width=5in]{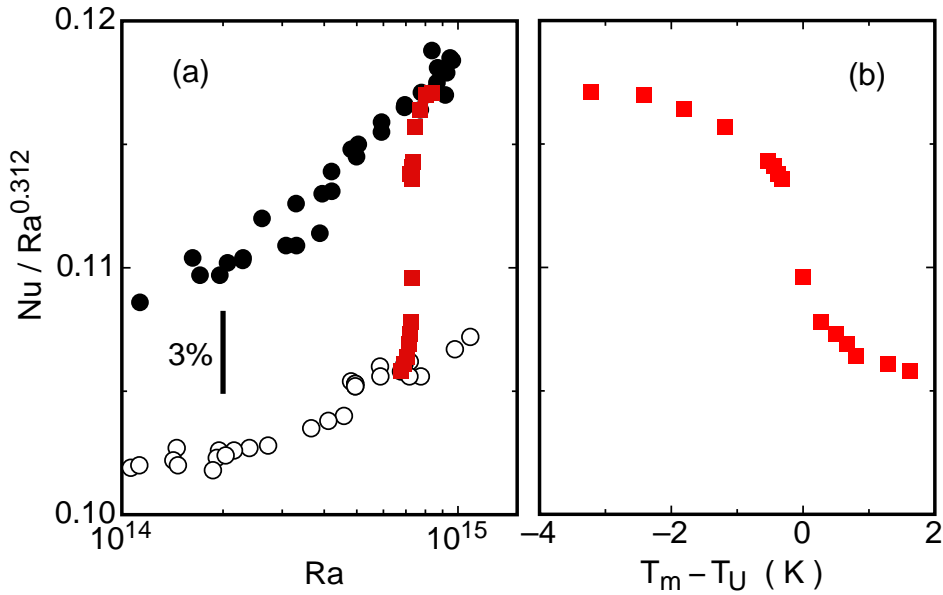}}
\caption{a): The reduced Nusselt number $\Nu/\Ra^{0.312}$ on a linear scale as a function of \Ra\ on a logarithmic scale. Solid (open) black circles: $T_m - T_U < -2$ K ($T_m - T_U > 2$ K). Solid red squares: $\Nu_{red} = \Nu/Ra^{0.312}$ at nearly constant $\Delta T \simeq 10.3$ K. b): $\Nu/\Ra^{0.312}$ at nearly constant $\Delta T \simeq 10.3$ K as a function of $T_m - T_U$ on linear scales.}
\label{fig:Nured_of_Ra_condtDT}
\end{figure}

In order to explore the dependence of the data on $T_m - T_U$ in more detail we show in Fig.~\ref{fig:Nured_of_Ra_condtDT}a the results for the reduced Nusselt number $\Nu_{red} = Nu/Ra^{0.312}$ on a linear scale as a function of \Ra\ on a logarithmic scale. Here it becomes apparent that the  $T_m - T_U < -2$ K data (solid circles) are higher than the $T_m - T_U > 2$ K data (open circles) by about 6\% near $\Ra = 10^{14}$ and about 10\% near $\Ra = 10^{15}$. In the classical regime $\Ra \alt 10^{13}$ (see Fig.~\ref{fig:hires} and Table~\ref{tab:clasicalfitparas}) the difference is 3.8\%. In Fig.~\ref{fig:Nured_of_Ra_condtDT}a we added a third set of data taken at nearly constant $\Delta T \simeq 10.3$ K but at various values of $T_m - T_U$. During these measurements $T_U$ was not controlled and determined by the balance between the heat input to the Uboot from the HPCF-II and the cooling to the surrounding laboratory. It varied over the narrow range from 24.3 to 25.5$^\circ$C. The sample temperature  $T_m$ was controlled by feedback loops and was changed in small steps from 21 to 27$^\circ$C. Since $T_m$ (and thus the fluid properties) changed, the data at constant $\Delta T$ led to a small variation of \Ra. One sees that they cover the $\Nu_{red}$ range from the upper to the lower branch.

The same constant $\Delta T$ results are shown also in Fig.~\ref{fig:Nured_of_Ra_condtDT}b, but as a function of $T_m - T_U$. Here one sees that the data become independent of $T_m - T_U$ when $|T_m - T_U| \agt 2$K. This is the reason why the majority of data (the open and solid black circles) were taken as a function of \Ra\ with $|T_m - T_U| \agt 2$ K.

\begin{figure}
\centerline{\includegraphics[width=5in]{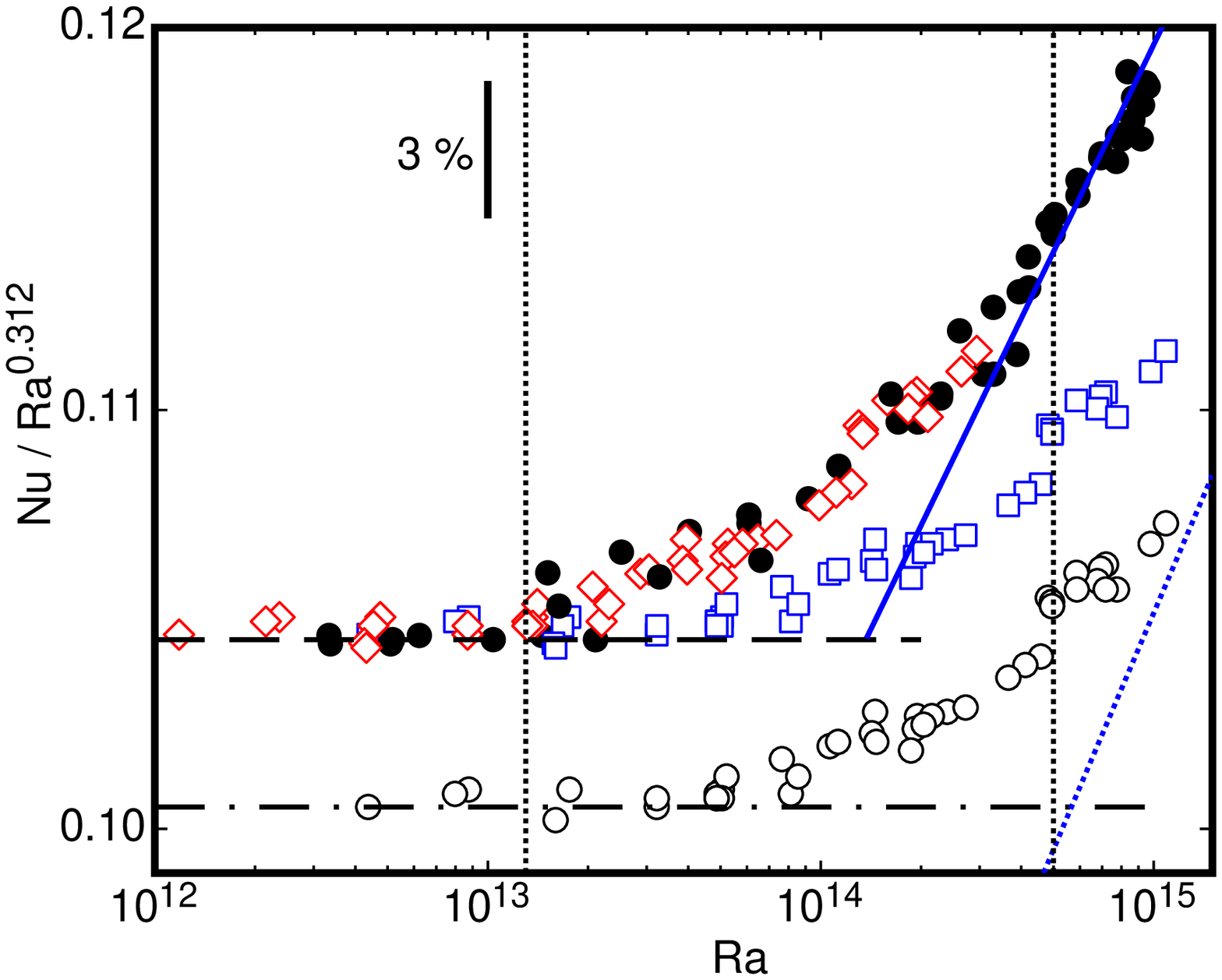}}
\caption{The reduced Nusselt number $\Nu/\Ra^{0.312}$ as a function of \Ra\ for the ``closed" sample. Black solid circles: $T_m - T_U \alt -2$K. Black open circles: $T_m - T_U \agt +2$K. Open squares (blue online): \Nu\ values of the open black circles multiplied by 1.04. Open diamonds (red online): \Ra\ values of the open squares divided by 3.7. The horizontal lines represent power laws for \Nu\ with $\gamma_{eff} = 0.312$ and $N_0 = 0.1005$ and 0.1044. 
The solid (dotted) line (blue online) through (near) the data at the largest \Ra\ corresponds to $\gamma_{eff} = 0.37$. 
}
\label{fig:Nured_of_Ra_closed}
\end{figure}

In Fig.~\ref{fig:Nured_of_Ra_closed} we show all the data for the two states with $|T_m - T_U| \agt 2$ K over the entire \Ra\ range accessible in the experiment. One sees the classical state with $\gamma_{eff} = 0.312$ for $\Ra \alt 10^{13}$.  At larger \Ra\ the two data sets trace out curves with remarkably similar shapes, albeit displaced both vertically and horizontally. To further explore the similarity between the two sets we  multiplied the open circles by 1.04 . This yielded the open blue squares, which now agree with the solid black circles in the classical range. Further dividing the \Ra\ values of the open blue squares by 3.7 yielded the open red diamonds. One sees that these two transformations yielded agreement within the experimental scatter between the data at large and small $T_m-T_U$. This shows that the shapes of the curves traced out by the two data sets are the same. 

Both data sets reveal a departure from the classical effective power law, with \Nu\  increasing more rapidly with \Ra\ than $\Ra^{0.312}$ when $\Ra > \Ra^*_1$ where $\Ra^*_1 \simeq 1.5\times 10^{13}$ for the solid circles and $\simeq 5\times 10^{13}$ for the open circles. Henceforth we shall concentrate on the results for $T_m - T_U < -2$ K. They continue to increase beyond the classical-state values, with an effective exponent that gradually becomes larger until $\Ra^*_2 \simeq 5\times 10^{14}$ is reached. Beyond  $\Ra^*_2$ one has $\gamma_{eff} \simeq 0.37\pm 0.01$ as shown by the blue solid line in the figure. This result is consistent with the prediction of an asymptotic exponent $\gamma = 1/2$ modified by logarithmic corrections in the ultimate state with turbulent boundary layers above the bottom and below the top plate. The recent prediction by Grossmann and Lohse  \cite{GL11} for the form of the logarithmic corrections differs from that given by Kraichnan \cite{Kr62}; but our data can not distinguish between these two theoretical results which both yield values of $\gamma_{eff}$ which are roughly in the range from 0.38 to 0.40.

It is worth noting that the data in the transition range $\Ra^*_1 < \Ra < \Ra^*_2$ have significantly greater scatter than the data in the classical regime $\Ra < \Ra^*_1$ or those in the ultimate regime $\Ra > \Ra^*_2$.  This indicates the existence of multiple states, presumably with subtly different BL configurations, during the complex transition from laminar to turbulent BLs.

We call attention to the fact that the transition range between the classical and the ultimate state can also be found, between about the same values of $\Ra_1^*$ and $\Ra_2^*$,  in results for the Reynolds number \cite{HFNBA12} and in measurements of vertical logarithmic temperature profiles that extend over much of the sample height \cite{ABGHLSV12}.

\subsection{Comparison with previous results.}
\label{sec:compare}

\subsubsection{Comparison with the Grenoble, Oregon, and DNS data.}

\begin{figure}
\centerline{\includegraphics[width=5in]{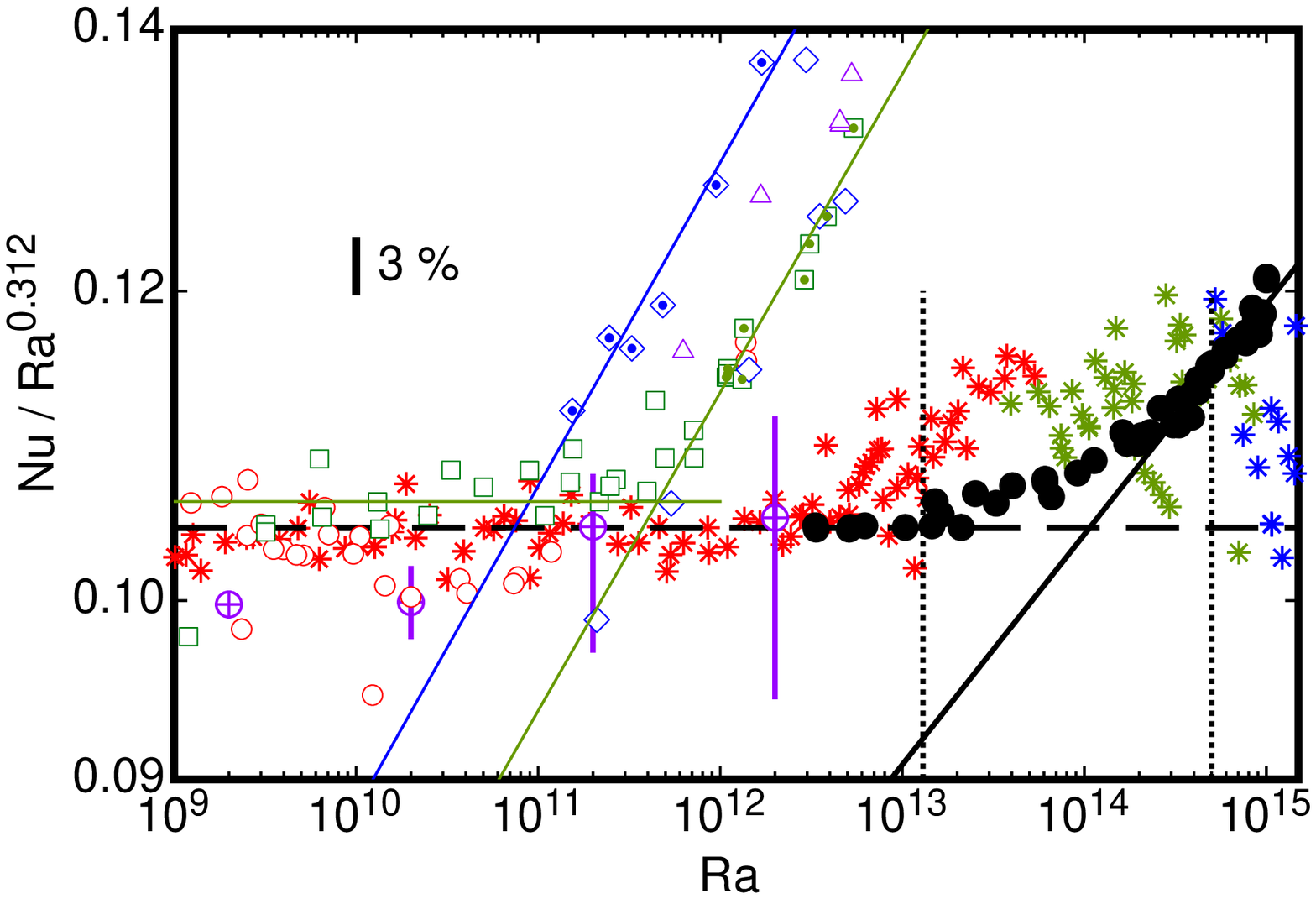}}
\caption{Comparison of the present results (solid black circles) with the Oregon (stars) \cite{NSSD00}, the Grenoble \cite{CCCCH01} (small open symbols), and the direct numerical simulation \cite{SVL10} (DNS) data. For the Oregon and Grenoble data we used different colors for different ranges of \Pra. Red: $\Pra < 1$. Green: $1 < \Pra < 2$. Blue: $2 < \Pra < 4$. Purple: $4 < \Pra < 8$. The DNS data are for $\Pra = 0.7$ and are shown as purple circles with plusses and error bars. The slanting green and blue line are power-law fits to the points with small dots in their centers.}
\label{fig:Nred_of_Ra_Oregon}
\end{figure}

\begin{figure}
\centerline{\includegraphics[width=5in]{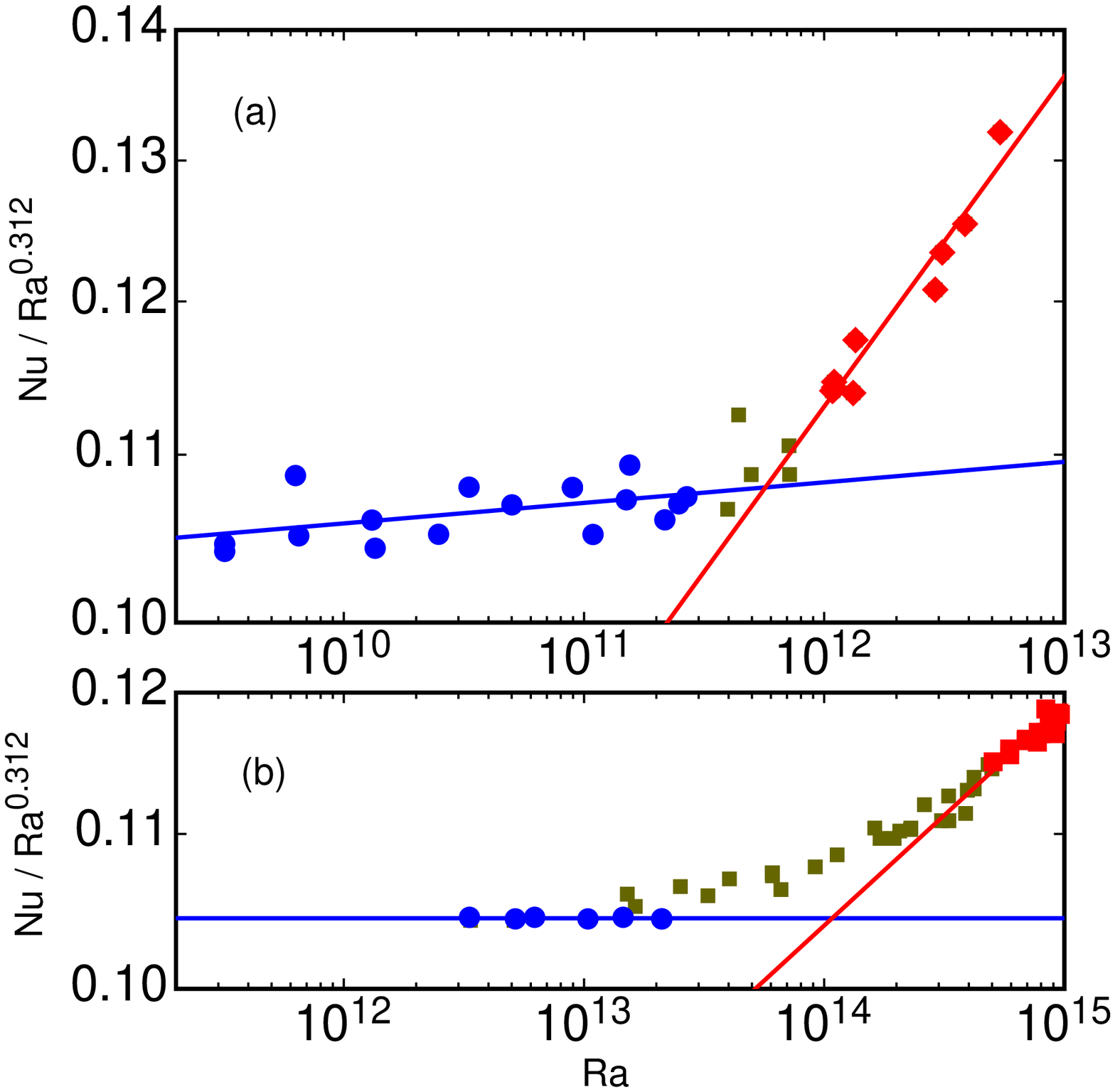}}
\caption{(a): A higher-resolution graph, on double logarithmic scales, of the Grenoble data \cite{CCCCH01} for $\Pra = 1.1$ and 1.3. A power-law fit to the blue circles (red diamonds) yielded $\gamma_{eff} = 0.317 \pm 0.003$ ($\gamma_{eff} = 0.382 \pm 0.006$). The green squares were not used in either fit. (b): Our data are plotted with the same resolution as that used in (a), but with the horizontal axis shifted by two decades.  A power-law fit to the blue circles (red diamonds) yielded $\gamma_{eff} = 0.3121 \pm 0.0004$ ($\gamma_{eff} = 0.371 \pm 0.01$). Our data show a transition region with a width of about 1.5 decades (green squares), whereas the Grenoble data suggest a sharp transition.}
\label{fig:Grenoble_detail}
\end{figure}

\begin{figure}
\centerline{\includegraphics[width=4in]{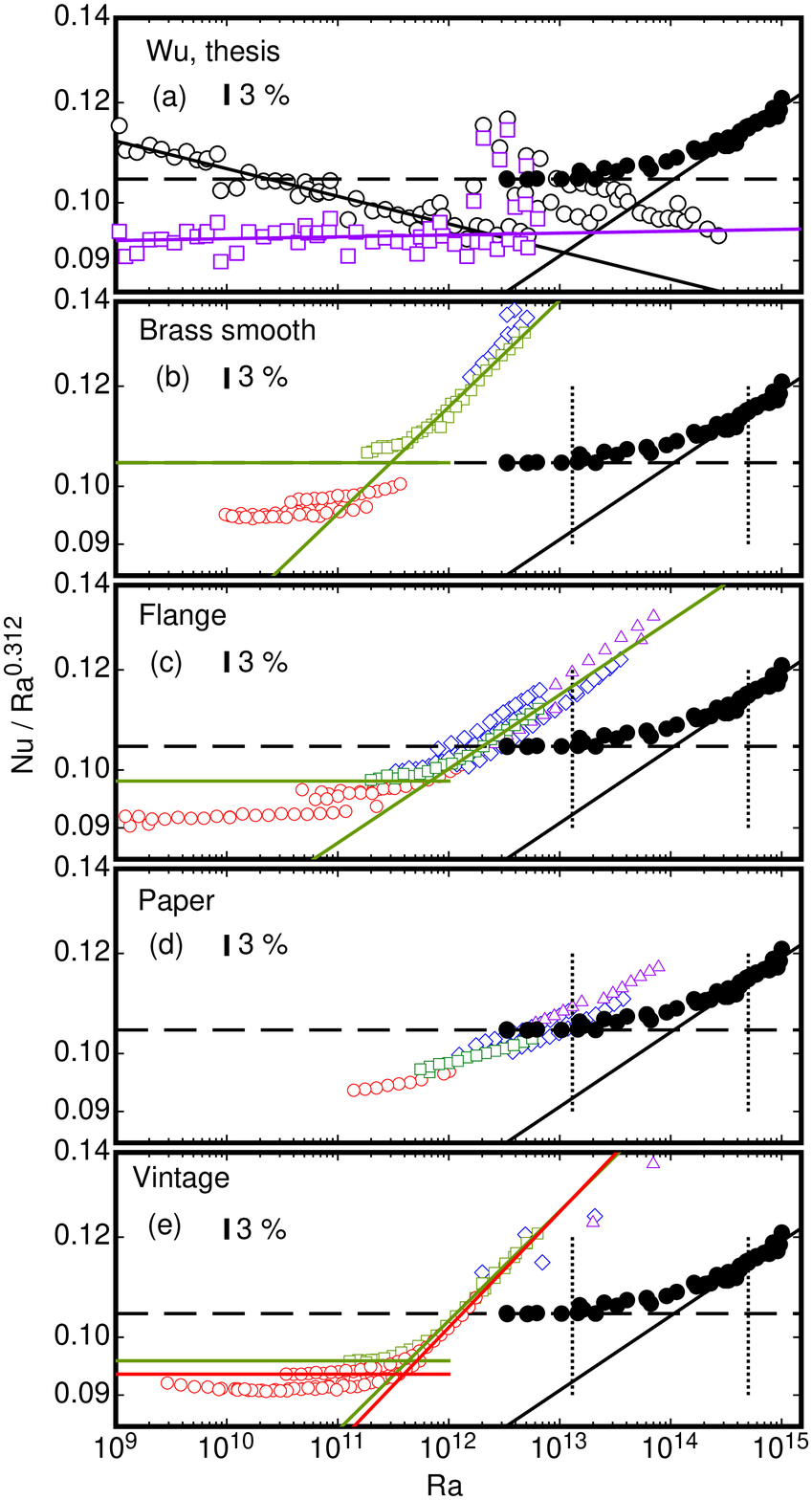}}
\caption{Comparison of the present results (solid black circles) with (a) the Chicago data \cite{Wu91}, (b) data from \cite{RGCH05}, and (c) to (e) data from \cite{RGKS10}. We used different symbols for different ranges of \Pra. Red open circles: $\Pra < 1$. Green open squares: $1 < \Pra < 2$. Blue open diamonds: $2 < \Pra < 4$. Purple open triangles: $4 < \Pra < 8$. The horizontal black dashed line is our best estimate for the classical state. The other lines are fits described in the text.}
\label{fig:Nred_of_Ra_Roche}
\end{figure}

Throughout this comparison we shall focus on our data for $T_m - T_U \alt -2$K; they will be shown as black solid circles in Figs.~\ref{fig:Nred_of_Ra_Oregon} and \ref{fig:Nred_of_Ra_Roche}. We mention again that, for our results, \Pra\  changed only over the narrow range from 0.79 near $\Ra = 3\times10^{12}$ to 0.86 near $\Ra = 10^{15}$. Throughout this comparison we shall show the data of others in red for $\Pra < 1.0$, green for $1 < \Pra < 2$, blue for $2 < \Pra < 4$, and purple for $4 < \Pra < 8$.

	In Fig.~\ref{fig:Nred_of_Ra_Oregon} our measurements are compared with those of Niemela et al. \cite{NSSD00} (Oregon data) which are given as stars. For the large range $10^9 \alt \Ra \alt 2\times 10^{12}$ those results  agree well with  the power-law fit to our data in the classical regime which gave $\Nu = 0.404 \Ra^{0.312}$. One can argue that they rise above  this power law as \Ra\ approaches $10^{13}$, perhaps signaling the beginning of a transition to the ultimate state, but in view of the scatter of the data this argument might not be convincing. It is noteworthy that the rise occurs when \Pra\ is still less that one. However, as \Ra\ increases beyond $10^{13}$, no further increase above the classical power law occurs. It may be that this saturation is due to an increase of \Pra, which begins to occur just in this \Ra\ range and which is expected to shift $\Ra^*$ to higher values. In summary, the Oregon data show a departure from the classical power law near $\Ra = 10^{13}$, but the evidence for having entered the ultimate state in our view remains inconclusive. We note that the original authors of this work believed that their data could be represented within their scatter and accuracy by a single power law with $\gamma_{eff} = 0.32$ over the entire \Ra\ range up to $10^{17}$, thus providing no evidence for an ultimate-state transition.

Also shown in Fig.~\ref{fig:Nred_of_Ra_Oregon}, as small open symbols, are the Grenoble data \cite{CCCCH01} that fall in the range of the figure. One can see that they form distinct groups, depending on \Pra. For $1 < \Pra < 2$ (green symbols) there are data in the classical regime. They yield the effective exponent of $0.317 \pm 0.003$, which is consistent with $\gamma_{eff} = 0.312$ as adopted by us when possible systematic errors, particularly due to uncertainties in the side-wall corrections \cite{Ah00,RCCHS01}, are taken into consideration. The actual values of $N_0 = \Nu/\Ra^{0.312}$ are also quite close to our result of 0.104. An average of the data for $10^9 < \Ra < 10^{11}$ is only 2\% higher, as shown by the horizontal green line in the figure. 

For $1 < \Pra < 2$  there is a sharp transition to a \Ra\ range where the effective exponent is larger than the classical value of 0.312.
In order to determine the Rayleigh number $\Ra_t$ at the transition and the effective exponent above it objectively, we selected a subset of points which we deemed to be above $\Ra_t$ and which seemed consistent with an effective power law. These points are indicated by a small dot in the open symbols. A power-law fit to these data and its intersection with the green horizontal line in the figure gave the parameters listed in Table~\ref{tab:param}. For $2 < \Pra < 4$ there also are sufficient data to warrant a power-law fit, but there are no classical-state data. We carried out the same analysis as for $1 < \Pra < 2$, and used the result for the classical state for $1 < \Pra < 2$ as the baseline to determine $\Ra_t$.
One sees from the Table that the exponents are, within their statistical errors, consistent with 0.38, a value deemed typical of the ultimate state. The values of $\Ra_t$, however, in our view are too low to correspond to the shear-induced boundary-layer transition to turbulence that is expected to be characteristic of the ultimate-state transition. Further, as noted in Ref.~\cite{RGKS10}, we see that  the larger \Pra\ value yielded a lower value of $\Ra_t$, which is opposite to expectations for the BL shear instability. However, this trend is not confirmed by the data for the larger \Pra\ range $4 < \Pra < 8$ (purple open triangles), which do not have enough points to warrant an independent power-law fit but which are seen to fall between the other two data sets. Thus we must conclude that the Grenoble data do not establish an unambiguous trend of $\Ra_t$ with \Pra. An explanation of the different values of $\Ra_t$ for the different data sets which seems likely to us is systematic errors of the equation of state or the transport properties near the critical point of helium that were used to calculate \Ra\ and \Nu; these errors could well change as the pressure and $T_m$ (and thus \Pra) are changed. 

The Grenoble data for $\Pra = 1.1$ and 1.3 are of sufficiently high precision and sufficiently detailed and plentiful to warrant a closer examination, as is done in Fig.~\ref{fig:Grenoble_detail} (a). There they are compared with our results, which are shown in Fig.~\ref{fig:Grenoble_detail} (b) on vertical and horizontal scales with the same resolution as in (a) but with the horizontal axis shifted by two decades. A remarkable difference between the two data sets is that the Grenoble data reveal a {\it sharp} transition suggestive of a continuous, or supercritical, bifurcation, whereas our data show a transition {\it range} of about 1.5 decades with indications of multi-stability in that range which is inconsistent with a continuous transition. We believe that a transition to a turbulent BL is unlikely to be sharp and continuous for at least two reasons. First, even for a uniform laminar BL 
the transition to turbulence does not occur at a unique value of the applied stress but rather will depend on the particular prevailing perturbations. In the time average this should lead to some rounding of the observed transition. Second, in the RB case the laminar BLs are  not uniform. Rather, due to plume emission, they are highly fluctuating systems. In addition,  they are non-uniform on longer length scales in the horizontal plane \cite{LX98} when a LSC is present. Because of their spatial inhomogeneity they are unlikely to undergo a simultaneous transition from the laminar to the turbulent state at all lateral positions. These inhomogeneities are consistent with the existence of a transition {\it range}, as observed by us.

Finally, in Fig.~\ref{fig:Nred_of_Ra_Oregon}, we show as purple open circles with plusses and error bars the results obtained from direct numerical simulation by Stevens et al. \cite{SVL10}. These data are for $\Ra \leq 2\times 10^{12}$. They do not show a transition for $\Ra \leq 2\times 10^{12}$ to a state with a larger effective exponent, and in that sense differ from the Grenoble data but agree with the Oregon data and with our results.

\begin{table}
\begin{center}
\begin{tabular}{cccccccc}
Sample                        & Grenoble                  & Grenoble               & Flange                   & Vintage               & Vintage                 &  Brass                     & HPCF-IIe \\
Ref.                        & \cite{CCCCH01}                  & \cite{CCCCH01}               & \cite{RGKS10}                   & \cite{RGKS10}               & \cite{RGKS10}                 & \cite{RGCH05}                     &  this work\\
\hline
$\Ra_{min}/10^{11}$                &$1.5$&$10$ & $20$  & $20$ &   $8$  & $2$  & $5000$  \\
$ \Ra_{max}/10^{11}$               &$17$&$60$ &$20$ &  $20$ &   $18$ & $20$  & $10000$  \\
$ \Pra_{min}$              &       3.4           &    1.0          & 1.72                         & 1.31         & 0.97                          & 1.74                         & 0.8 \\
$ \Pra_{max} $            &        3.7                    &     1.3                   &1.72                         & 1.73                        & 0.97                          & 1.74                          & 0.8 \\
$ \Nu_{red}^{cl}$&0.106                    &        0.106            & 0.098                      & 0.096                     & 0.094                       &0.104                         & 0.104          \\
$ \gamma^{eff}$          &0.395                        &     0.382              &0.371                        & 0.399                     & 0.404                        & 0.396                       & $0.37\pm 0.01$ \\
$ \Ra_{t}/10^{11}$                  &$0.9$   & $5$ & $7$  & $4$  & $4$    & $2$   & $1100$   \\
\end{tabular}
\end{center}
\caption{Parameters obtained from power-law fits to various data sets. $\Ra_{min},\Ra_{max}, \Pra_{min}$, and $\Pra_{max}$ are the limits of the data sets used in the power-law fits. $ \Nu_{red}^{cl}$ is the reduced Nusselt number $\Nu/\Ra^{0.312}$ used in the classical regime to determine the transition Rayleigh number $\Ra_t$, which is taken to be the intercept with the power-law fit above $\Ra_t$.}
\label{tab:param}
\end{table}

\subsubsection{Comparison with the Chicago data: }

The ``Chicago" data are the earliest measurements of \Nu\ at very large \Ra. For $\Gamma = 0.5$ they were reported by X. Z. Wu \cite{Wu91}. These results, after a correction \cite{Ni_reanal} which shifted the data upward by an amount that varied smoothly from about 6 \% near $\Ra = 10^9$ to about 10\% near $\Ra = 10^{14}$, are shown in Fig.~\ref{fig:Nred_of_Ra_Roche} (a) as open black circles. Their trend with \Ra\ suggests $\gamma_{eff} \simeq 0.29$ \cite{Ah00}. However, they had not been corrected for the finite sample-wall conduction \cite{Ah00,RCCHS01}. That correction, based on Model 2 of \cite{Ah00}, was applied in Ref.~\cite{Ah00} and yielded the open purple squares in the figure. Those corrected data are consistent with $\gamma_{eff} = 0.312$, but the pre-factor $N_0$ of a power-law fit is about 10 \% lower than indicated by our results.

Just above $\Ra = 10^{12}$ there is a discontinuity in the data, but this apparent ``transition" is not to a state with a larger $\gamma_{eff}$.  We believe that this phenomenon is associated with a change of the place in the phase diagram near the critical point of helium where the data were taken, and that it is due to errors in the equation of state rather than a genuine change in the dependence of \Nu\ on \Ra. Thus we conclude that the Chicago data do not reveal any evidence of a transition to the ultimate state below their largest $\Ra \simeq 3\times 10^{14}$. Given the increase of \Pra\ with \Ra\ for these data and the expected dependence of $\Ra^*$ on \Pra, we find that the absence of a transition to the ultimate state in these data is consistent with the Oregon data and with our results. 

\subsubsection{Comparison with data from Roche {\it et al.}.}

Finally, in Fig.~\ref{fig:Nred_of_Ra_Roche}(b) to (e), we display several data sets obtained in different sample cells, all with $\Gamma = 0.5$, by Roche and co-workers \cite{RGCH05,RGKS10}. Many of them show a transition at $\Ra_t$ from the classical state to a state with a larger $\gamma_{eff}$. Whenever there are   adequate classical-state data, these sets are consistent with the exponent 0.312, but the pre-factor of the corresponding power law varied between different sets and generally was somewhat lower than our result $N_0 = 0.104$.  We analyzed the sets with sufficient data both above and below $\Ra_t$ as described above for the Grenoble data. The results are given in Table~\ref{tab:param}. The effective exponent varies from 0.371 to 0.404. A reason for this variation is not obvious to us. The value of $\Ra_t$ varied significantly as well, from $0.9\times 10^{11}$ to $7\times 10^{11}$, again for non-obvious reasons. 

\begin{figure}
\centerline{\includegraphics[width=4in]{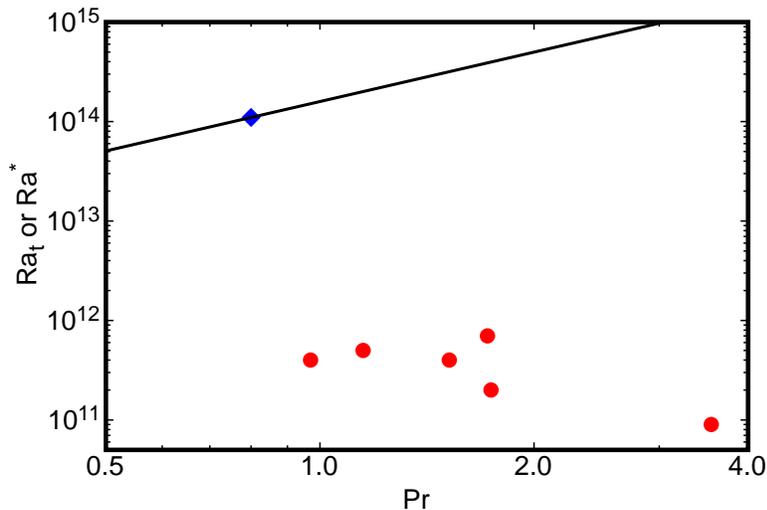}}
\caption{The transition Rayleigh number $\Ra_t$ as a function of \Pra. The data are from Table~\ref{tab:param}. Blue diamond: HPCF-IIe. Red circles: The remaining points in the table. The solid line is a fit of the theoretically expected dependence $\Ra_t \propto \Pra^{1.65}$ (Eq.~\ref{eq:Ra*}) to the HPCF-IIe point.}
\label{fig:Ra*_of_Pr}
\end{figure}

In Fig.~\ref{fig:Ra*_of_Pr} we show $\Ra_t$ as a function or \Pra\ as red solid circles. A trend with \Pra\ is, in our view, not firmly established, although the point at the largest \Pra\  suggests a decrease with increasing \Pra. Such a decrease would be contrary to the theoretical expectation for a shear-driven BL transition to turbulence. The theoretically expected result is obtained by assuming that the transition to the ultimate state occurs at a Rayleigh number $\Ra^*$ at which the bulk Reynolds number (and thus also the shear Reynolds number $\Rey_s^* \propto \sqrt{\Rey^*})$) attains a certain constant value $\Rey^*$. Taking $\Rey \propto \Ra^{0.423}/\Pra^{0.70}$, one finds 
\be
\Ra^* = 1.6\times 10^{14} \Pra^{1.65}\ .
\label{eq:Ra*}
\ee
 Here the constant coefficient was adjusted so that the relation passes through the point $\Ra^* = 1.1\times 10^{14}$ for $\Pra = 0.8$ shown as the blue diamond in the figure, which is the result from the present work. The line through that point corresponds to Eq.~\ref{eq:Ra*}. One sees that the red circles do not reproduce the predicted increase of $\Ra^*$ with \Pra, as already noted by Roche et al. \cite{RGKS10}. However, we do not find convincing evidence in these data that larger values of \Pra\  {\it favor} the transition to the state with the larger exponent.

\subsection{Open sample}
\label{sec:open}

\begin{figure}
\centerline{\includegraphics[width=4in]{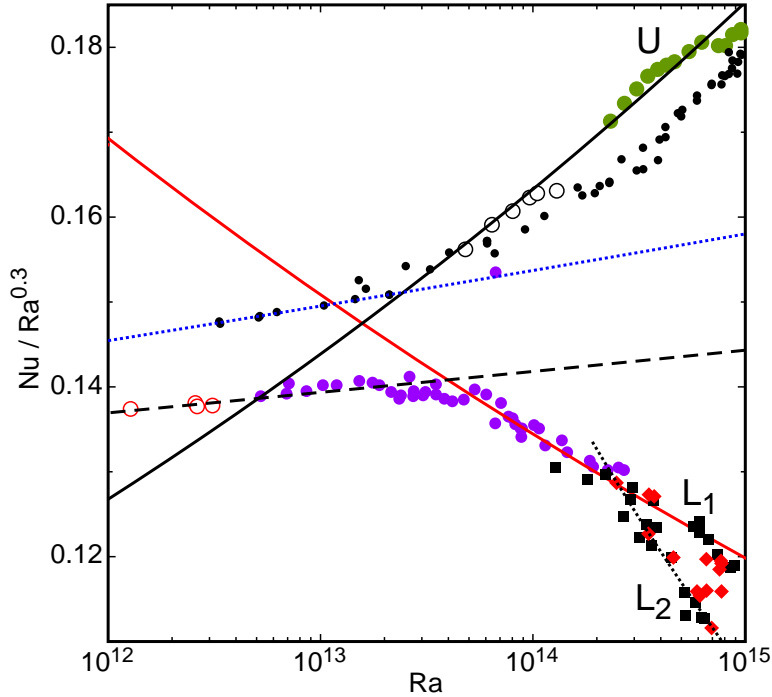}}
\caption{The reduced Nusselt number $\Nu/\Ra^{0.3}$ as a function of the Rayleigh number $\Ra$ on a logarithmic scale. The purple solid circles, black solid squares, and red solid diamonds are from HPCF-I, HPCF-IIa, and HPCF-IIb respectively, using SF$_6$ at various pressures. The open red circles are from HPCF-I using N$_2$. 
All of those data were shown before in Ref. \cite{AFB09}. The open black and solid green circles are new data from HPCF-IIb using SF$_6$ at 12.2 bars ($Pr = 0.823$) and 18.8 bars ($Pr = 0.863$) respectively. The small black dots are the present results for HPCF-IIe (the closed sample). The lines correspond to power laws $\Nu = N_0 \Ra^{\gamma_{eff}}$ with $N_0 = 0.111$, $\gamma_{eff} = 0.308$ (dashed black line), $N_0 = 0.674, \gamma_{eff} = 0.25$ (red solid line), $N_0 = 12.3, \gamma_{eff} = 0.16$ (black dotted line), $N_0 = 0.0277, \gamma_{eff} = 0.36$ (black solid line), and $N_0 = 0.104, \gamma_{eff} = 0.312$ (dotted blue line). The labels U, L$_1$, and L$_2$ identify the "upper branch", "lower branch 1", and "lower branch 2".
}
\label{fig:Nured_of_Ra_open}
\end{figure}

As discussed in Sec.~\ref{sec:apparatus}, initially (for HPCF-I, IIa, and IIb) gaps with an average width of about one mm were left between the top and bottom plates and the side wall in order to permit the SF$_6$ to pass from the Uboot into the sample. These samples are referred to as the ``open" samples. Although the gap widths were negligible compared to the sample dimensions, it was appreciated later that they could significantly modify the system, depending on the temperature difference $T_m - T_U$ between the sample ($T_m$) and the Uboot ($T_U$). When $T_m < T_U$ ($T_m > T_U$), then the sample density  is larger (smaller) than the density of the fluid in the Uboot. In the presence of gravity this density difference will lead to flow through the gaps, in the upward (downward) direction when $T_m - T_U > 0~(< 0)$. Such an externally imposed  flow can be expected to modify the ideal RBC system.    

Previously \cite{FBA09,ABFH09,AFB09} we presented some measurements of \Nu\ 
for the open system. These measurements were made with $T_m = 25^\circ$C. During those early stages of this investigation $T_U$ was not actually measured, but more recent experience suggests that it was between 23 and 24$^\circ$C. Thus the data are for $T_m - T_U > 0$. The results are shown again in Fig.~\ref{fig:Nured_of_Ra_open} in the form of $\Nu/\Ra^{0.3}$ as a function of $\Ra$.  The purple solid circles, black solid squares, and red solid diamonds are from HPCF-I, HPCF-IIa, and HPCF-IIb respectively and were obtained with SF$_6$. The open red circles were measured  using Nitrogen in HPCF-I.  For comparison the closed-sample data are shown as small black dots, with the power-law fit in the classical range (see Fig.~\ref{fig:hires}) with $\gamma_{eff} = 0.312$ shown as a dotted blue line. For $\Ra \alt 4\times 10^{13}$ the open-sample data are described well by a power law with $\gamma_{eff} = 0.308$ (the dashed black line), reasonably consistent with the classical state; however, they are somewhat lower than the closed-sample data.

With increasing \Ra\  the measurements from HPCF-I (purple solid circles in Fig.~\ref{fig:Nured_of_Ra_open}) revealed a transition to a new state beyond $\Ra_{t1} \simeq 4\times 10^{13}$. Within our resolution \Nu\ was continuous at that transition, and the transition was sharp. In those respects the transition differed from the ultimate-state transition seen for the closed sample but was similar to the transitions at $\Ra_t = {\cal O}(10^{11})$ found by Chavanne et al. \cite{CCCCH01} (see Fig.~\ref{fig:Nred_of_Ra_Oregon} and \ref{fig:Grenoble_detail}(a)) and Roche {\it et al.} \cite{RGKS10}. However, the present case is very different from the previous ones in that $\gamma_{eff}$ decreased whereas  for the prior work it increases. Above $\Ra_{t1}$ the data could be described well by a power law with $\gamma_{eff} = 0.25$, as shown by the solid red line in the figure. We shall refer to this state as the ``$L_1$" state or branch. 

In a recent paper \cite{GL11} Grossmann and Lohse pointed out that, with increasing \Ra,  the transition away from the classical state can take several possible forms. Whereas previously \cite{Kr62,Sp71} the ultimate-state transition was assumed to involve a simultaneous shear-induced transition to turbulence in both the viscous and the thermal BL, this need not be the actual sequence of events. They proposed three possibilities that may prevail when the viscous BL becomes turbulent. One of these is that the thermal BL remains laminar ({\it i.e.} of the Prandtl-Blasius type in the time average) and that the heat transport is background dominated (see Sec. III.B of \cite{GL11}). For that case they derive $\Nu \sim \Ra^{1/5}$, with logarithmic corrections which yield $\gamma_{eff} \simeq 0.22$ to 0.23. Within experimental and theoretical uncertainties this is consistent with our result  $\gamma_{eff} \simeq 0.25$ for the $L_1$ branch.

Further measurements, with HPCF-IIa (black solid squares in Fig.~\ref{fig:Nured_of_Ra_open}) and HPCF-IIb (red solid diamonds in Fig.~\ref{fig:Nured_of_Ra_open}), revealed the existence of yet another branch which we labeled as $L_2$. The transition to this branch occurred at $\Ra_{t2} \simeq 2.5\times 10^{14}$. The new branch co-existed with $L_1$. The precise conditions that determined which branch was chosen by the system were not explored in detail, but they involved $T_m - T_U$ and thus the strength of the presumed external current entering or leaving the sample. A power-law fit to the $L_2$ data yielded $\gamma_{eff} \simeq 0.16$. This result is consistent with the theoretical value (see Sec. III.A of \cite{GL11}) $\gamma =   1/8$ which logarithmic corrections for a state with a turbulent viscous BL and a laminar thermal BL, with the heat transport dominated by the emission of plumes. For this state the logarithms modify $\gamma$ so that $\gamma_{eff} \simeq 0.14$, not very different from the experimental value.

When it was realized that $T_m - T_U$ played an important role in the choice of the state of the system, new measurements were made with HPCF-IIb and with $T_m - T_U \alt -2$ K. Those data are shown as open black and solid green circles in Fig.~\ref{fig:Nured_of_Ra_open}. They fall somewhat above the closed-sample data (small black dots), but seem reasonably consistent with a perhaps slightly modified transition to the ultimate state. This upper branch is labeled ``$U$". A power-law fit to these data yielded $\gamma_{eff} = 0.36$ and is represented by the solid black line in the figure.

\subsection{Half-open sample}
\label{sec:halfopen}
 
\begin{figure}
\centerline{\includegraphics[width=4in]{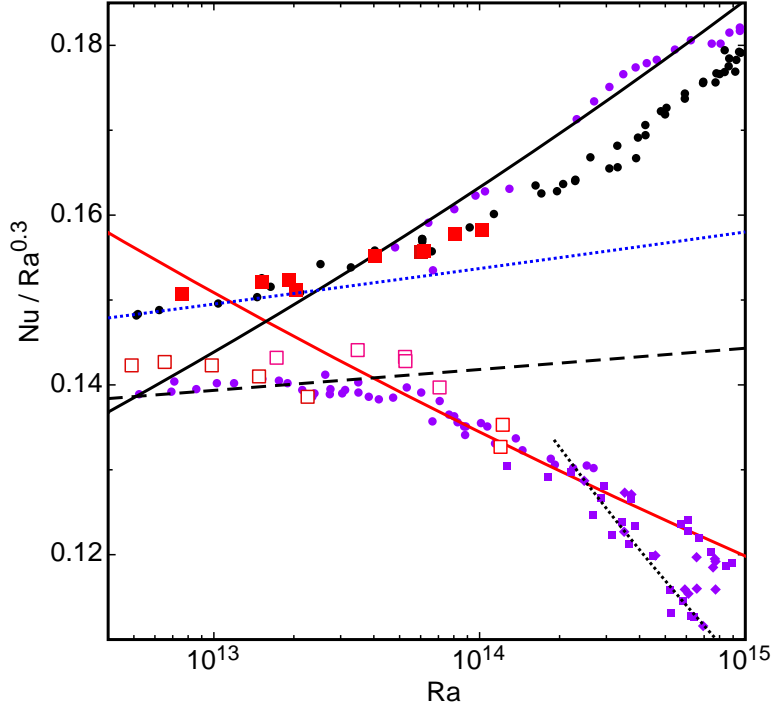}}
\caption{The reduced Nusselt number $\Nu/\Ra^{0.3}$ as a function of the Rayleigh number $\Ra$ on a logarithmic scale. Small purple symbols: open sample (see Fig.~\ref{fig:Nured_of_Ra_open}). Small black circles: closed sample, $T_m - T_U < -2$ K (see Fig.~\ref{fig:Nured_of_Ra_closed}). The larger red symbols are for the half-open sample. Solid squares:  $T_m - T_U < -2$ K. Open squares:  $T_m - T_U > 2$ K.}
\label{fig:Nured_of_Ra_halfopen}
\end{figure}

As a first attempt to prevent the external currents through the sample due to the  difference between $T_m$ and $T_U$, we sealed the bottom plate to the side wall but left the gap between the top plate and the side wall so that the gas could still enter the sample. We refer to this case, which is HPCF-IIc, as the half-open sample. The results for \Nu\ are shown in Fig.~\ref{fig:Nured_of_Ra_halfopen} as solid red squares for $T_m - T_U < -2$ K and as open red squares for $T_m - T_U > 2$ K. As might be expected, the data for $T_m - T_U < -2$ K agree well with the results from the closed sample (small solid black dots). This is so because the relatively dense gas, which in the open sample escapes through the gap at the bottom plate, can not do so in this case. However, the case where the sample gas is less dense than the gas in the Uboot ($T_m - T_U > 2$ K, open squares) seems to be influenced by external currents near the top plate and the corresponding data are close to those for the open sample (purple solid circles). Indeed they even reveal the transition at $\Ra_{t1}$ to the $L_1$ state.

 \section{Summary}
 \label{sec:summary}
 
In this paper we reported results for the Nusselt number \Nu\ over the range of the Rayleigh number \Ra\  from $3\times 10^{12}$ to $10^{15}$. Data were presented for three different sample cells, all of cylindrical shape and of aspect ratio $\Gamma = 0.50$. The cells, known as the High Pressure Convection Facilities (HPCFs), were located in a pressure vessel referred to as the ``Uboot" of G\"ottingen. The Uboot and the HPCF were filled with the gas sulfur hexafluoride (SF$_6$) at various pressures up to 19 bars. This fluid had a Prandtl number \Pra\ which varied over the narrow range from 0.79 to 0.86 as the pressure (and thus \Ra) changed from its smallest to its largest value. 

One of the samples was completely sealed. In that case a 2.5 cm diameter tube penetrated the side wall at mid height and allowed the SF$_6$ to enter the HPCF from the Uboot; that tube was then closed off by a remotely operated valve after an equilibration time of several hours and before measurements were made. This version of the HPCF is known as the ``closed" sample. Another version was allowed to have a small gap, of width approximately equal to one mm, between the side wall and the top and bottom plate to allow the sample gas to enter the HPCF. This version is known as the ``open" sample. A third version, known as the ``half-open" sample, had the side wall sealed to the bottom plate while the gap was allowed to persist at the top plate. 

It turned out that the three samples produced qualitatively different dependences of \Nu\ on \Ra. Only the closed sample can be regarded as corresponding well to the idealized closed-system Rayleigh-B\'enard problem. For the open and half-open samples gas currents were able to enter and leave the sample through the gaps at the plates, driven by the small density differences that existed between the sample fluid at a mean temperature $T_m$ and the fluid in the Uboot at a temperature $T_U$. The results for $\Nu(\Ra)$ were then qualitatively different depending on whether $T_m$ was larger or smaller than $T_U$. Nonetheless the open and half-open samples produced interesting effects. Under certain conditions there was a sharp but continuous transition  from a state below $\Ra_t \simeq 4\times 10^{13}$ where $\Nu \sim \Ra^{\gamma_{eff}}$ with $\gamma_{eff} \simeq 0.31$ to a state above it with $\gamma_{eff} \simeq 0.25$. Another state with $\gamma_{eff} \simeq 0.16$ was found as well. Possible explanations of these findings in terms of different laminar or turbulent conditions in the thermal and viscous boundary layers adjacent to the top and bottom plates were offered by Grossmann and Lohse \cite{GL11}. We refer the reader to Sects.~\ref{sec:open} and \ref{sec:halfopen} as well as to Ref.~\cite{GL11} for a more detailed discussion of these interesting phenomena.

The primary focus of this paper has been on the closed sample. Even for this case the results for $\Nu(\Ra)$ depended somewhat on $T_m - T_U$, but in contradistinction to the open and half-open samples the dependence was simply a shift of the curve without any change in shape. This dependence persisted in spite of the extensive shielding of the sample that was provided, and we do not know its origin. However, for $\Ra \alt 10^{13}$ we found that the effective exponent of \Nu\ was $0.312 \pm 0.002$, regardless of $T_m$ and $T_U$. This exponent value is consistent with many other measurements at smaller \Ra. It also agrees quite well with the value 0.323 with follows from a power-law fit to data generated from the Grossmann-Lohse model \cite{GL01} for $\Pra \simeq 0.8$ (albeit for $\Gamma = 1.00$) in the range $10^{12} \alt \Ra \alt 10^{13}$. Thus we believe that the state of the system with $\Ra \alt 10^{13}$ is the classical state of RBC with laminar boundary layers below the top and above the bottom plate.

In the range $\Ra_1^* \leq \Ra \leq \Ra_2^*$, with $\Ra_1^* \simeq 1.5\times 10^{13}$ and $\Ra_2^* \simeq 5\times 10^{14}$, the system gradually underwent a transition and $\gamma_{eff}$ increased from 0.32 to 0.37 as \Ra\ changed by about one and a half decades. We believe that this phenomenon reflects the transition to the ultimate state predicted by Kraichnan \cite{Kr62} and re-examined recently by Grossmann and Lohse \cite{GL11}. In the ultimate state, which we found for $\Ra > \Ra_2^*$, the effective exponent was $0.37 \pm 0.01$. This is reasonably consistent with the predicted asymptotic exponent $\gamma = 1/2$ and the expected logarithmic corrections \cite{GL11}. Our conclusion that the state above $\Ra_2^*$ is the ultimate state is supported by transitions and exponent-values found in simultaneous measurements of the Reynolds number \Rey\ \cite{HFNBA12} (a detailed discussion of those results is beyond the scope of this paper). Evidence for the transition range, with about the same values of $\Ra_{1,2}^*$,  can be found also in recent measurements of logarithmic vertical temperature profiles that extend throughout most of the sample \cite{ABGHLSV12}. 

It is worth emphasizing that the transition from the classical to the ultimate state is not a continuous transition where $\Nu(\Ra)$ is continuous but its effective exponent changes discontinuously such as was found at $\Ra_t \simeq 4\times 10^{13}$ in the open sample. Instead there was a transition range spanning a factor of 30 or so in \Ra\ over which the transition occurred. In this transition range the results for $\Nu(\Ra)$ scattered more than below or above it and were often irreproducible at the level of our resolution from one point to another, suggesting the existence of many states which differed in detail. 

The observed transition {\it range} (as opposed to a unique value of $\Ra^*$) is not surprising for two reasons. First, transitions involving shear-flow instabilities are known to depend on the size of prevailing local perturbations and thus will occur at different values of the \Rey\ (or in our case \Ra) number. In addition, in the present case the basic state, i.e. classical RBC, is known to have laminar boundary layers which are non-uniform in the horizontal plane \cite{LX98}. Since the BLs and the shear applied to them by the turbulent bulk are known to be spatially inhomogeneous, one should not expect a sharp transition.

In this paper we also provided a re-examination of earlier measurements. Some of these indicated  a transitions to a state where $\gamma_{eff}$ assumes a value near 0.38, but the transitions occurred at the rather low values of $\Ra_t$ between $10^{11}$ and $10^{12}$. All of these data involved a range of \Pra. We separated them into subsets corresponding to a unique value, or spanning only a small range, of \Pra. Within each subset one sees that the scatter of these data is much less than it is when all \Pra\ values are considered jointly. Some of the subsets reveal a well defined transition. These transitions differ qualitatively  from the one observed by us in that they are sharp and continuous, without any resolved transition range. Estimates of the shear Reynolds number for them are in the range near 100 or less, which to us seems too low for a turbulent boundary-layer shear transition. Based on our measurements of \Rey, and on recent calculations by DNS of the shear Reynolds number $\Rey_s$ \cite{WSW12}, as well as on earlier estimates \cite{GL02}, indicate that the boundary-layer shear transition at $\Ra^*$ to the ultimate state should not occur until $\Ra = {\cal O}(10^{14})$ is reached. Thus, in our view, the transitions observed near $\Ra = 10^{11}$ to $10^{12}$ are unrelated to the ultimate-state transition; but we can not offer an alternative explanation for their existence.

This paper concludes with a number of Appendices which examine the possible effect of several experimental factors on our results. None of them are found to have a significant influence.

\ack

We are very grateful to the Max-Planck-Society and the Volkswagen Stiftung, whose generous support made the establishment of the facility and the experiments possible. The work of G.A. was supported in part by the U.S National Science Foundation through Grant DMR07-02111. We thank Andreas Kopp, Artur Kubitzek, and Andreas Renner for their enthusiastic technical support. We are very grateful to Holger Nobach for many useful discussions and for his contributions to the assembly of the experiment.

\appendix

\section{Effect of tilting the sample}
\label{app:tilt}

All data for this study were taken with HPCF-IIe (the closed sample). Measurements were made with a leveled sample (a tilt angle $\beta < 0.1$ mrad, run 100918 to 110102), and with the same sample tilted relative to gravity. The tilt was used to localize the azimuthal orientation of the large-scale circulation plane. Based on measurements with smaller samples of aspect ratio $\Gamma = 1.00$ \cite{ABN06} and 0.50 \cite{WA12}, both with $Pr \simeq 4$, we do not expect a measurable influence of the tilt on the heat transport for our case of $\Gamma = 0.50$ and $\Pra \simeq 0.8$. This is indeed born out by the data.

\begin{figure}
\centerline{\includegraphics[width=6in]{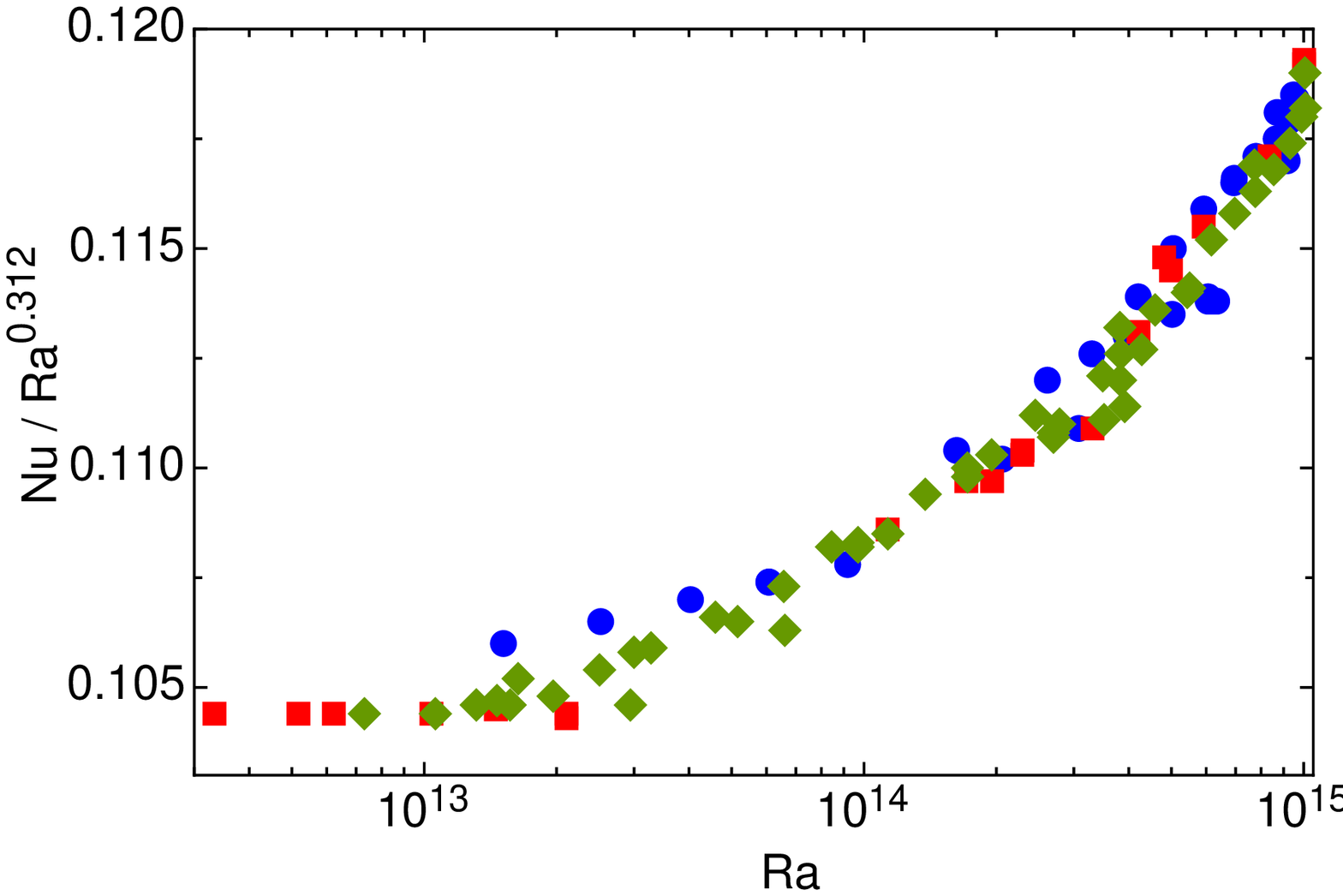}}
\caption{The reduced Nusselt number $\Nu/\Ra^{0.312}$ as a function of the Rayleigh number \Ra\ for $T_m - T_U < -2$K as a function of the Rayleigh number for three different tilt angles $\beta$ of the sample relative to gravity. Blue circles: ``level", $\beta < 10^{-4}$ rad. Green diamonds: $\beta \simeq 0.003$ rad. Red squares: $\beta \simeq 0.014$ rad.}
\label{fig:tilt}
\end{figure}

In one case the tilt angle was $\beta = 3$ mrad (run 110115 to 110610). The tilt direction was at an angle $\theta_\beta = 3\pi/2$ rad relative to the azimuthal origin of the sidewall thermometers that were used to monitor any large-scale circulation. In the other case we had $\beta = 14$ mrad and  $\theta_\beta = ?$ rad (run 110618 to 110919). The results for the Nusselt number for the data with $T_m - T_U < -2$K are shown in Fig.~\ref{fig:tilt} as blue circles (no tilt), green diamonds ($\beta = 3$ mrad) and red squares ($\beta = 14$ mrad). There is no significant difference between the three data sets, showing that the effect of the tilt on \Nu\ is well below one percent.  

\section{Non-Boussinesq effects}
\label{app:NOB}

\begin{figure}
\centerline{\includegraphics[width=5in]{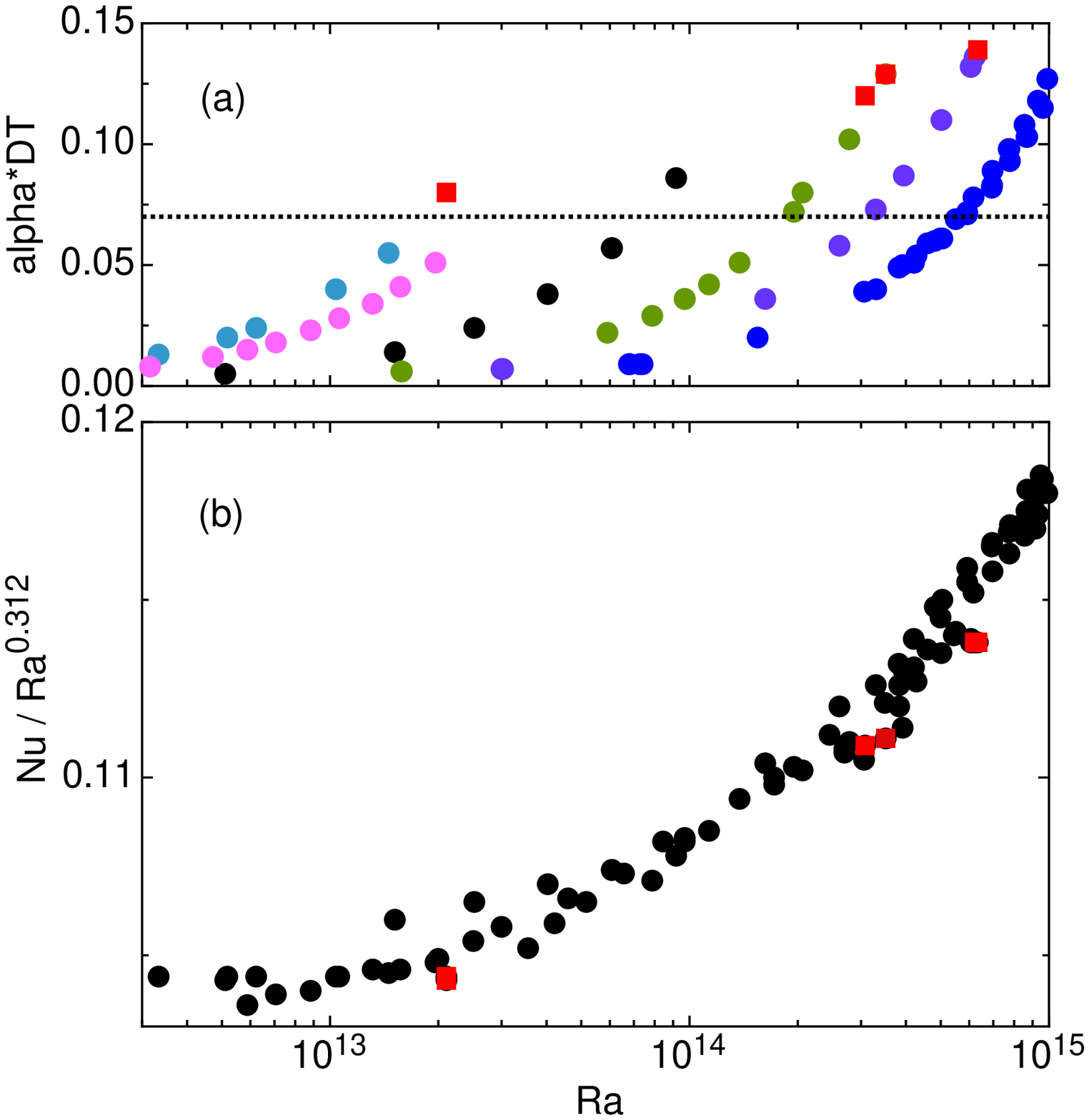}}
\caption{a): The parameter $\alpha \Delta T$ as a function of \Ra. The circles of various colors are, from left to right, for approximately 4, 5, 8, 12, 15, and 19 bars. The red squares are points for which $19 < \Delta T <21$ K. The dotted horizontal line indicates the perfect-gas value 0.071 for $\Delta T = 21.0$ K and $T_m = 294.2$ K.
b): The reduced Nusselt number for $T_m - T_U < -2$ K as a function of the Rayleigh number. Black circles: $\Delta T < 19$K. The red squares are points for which $19 < \Delta T < 21$K and correspond to the red squares in a).}
\label{fig:NOB}
\end{figure}

Non-Oberbeck-Boussinesq (NOB) effects \cite{Ob79,Bo03} on \Nu\  in a gas have been studied quantitatively using ethane gas as the fluid \cite{AFFGL07,ACFFGLS08}. Although variations of all fluid properties contribute in principle, for gases the primary contribution comes from the temperature dependence of the density and can be described approximately by the parameter $\alpha \Delta T$ (see Fig.~4 of \cite{AFFGL07}). All our measurements were made with $\Delta T \alt 21$K where we expect NOB effects to be small. For a perfect gas one has $\alpha \Delta T = \Delta T/T$ with $T$ in K, which for $\Delta T \simeq 20$K and at a mean temperature of 21$^\circ$C or 294 K is  close to 0.07. However, the properties of SF$_6$ in our pressure range up to 19 bars show large deviations from those of a perfect gas. Figure~\ref{fig:NOB}a shows $\alpha \Delta T$ as a function of \Ra\ for the data with $T_m - T_U < -2$ K. The different-colored circles are, from left to right, for different isobars with  pressures of  approximately 4, 5, 8, 12, 15, and 19 bars. For the points shown as red squares we had $19 < \Delta T < 21$ K, which were the largest temperature differences employed in this work. The dotted line is the approximate perfect gas result. One sees that $\alpha \Delta T$ was mostly below 0.1 and always below 0.14, but at the highest $\Delta T$ and pressure exceeded the perfect gas value by about a factor of two.

In Fig.~\ref{fig:NOB}b we show the reduced Nusselt number for the data with  $T_m - T_U < -2$ K.  All data are shown as solid black circles, except for the ones with $19 < \Delta T < 21$ K which are given as red squares. One sees that the points with large $\Delta T$ are systematically lower than the remaining results, but only by about 1.4 percent at the highest pressure where $\alpha \Delta T$ is largest and by about 0.7 percent at the lowest pressure where $\alpha \Delta T$ is about a factor of two smaller. It is surprising that NOB effects {\it reduce} \Nu\ in the present case because in the case of ethane \cite{AFFGL07,ACFFGLS08} they {\it enhanced} \Nu, albeit by only a small amount. A reduction of \Nu\ due to NOB effects was observed, however, in the case of a liquid, namely water \cite{ABFFGL06}, but again only by a percent or so for $\Delta T \simeq 20$ K. In any case, what matters for the present work is that NOB effects are negligible for nearly all of our data.

\section{The parameter $\xi$ of Niemela and Sreenivasan}
\label{app:NS_NOB}

Recently Niemela and Sreenivasan \cite{NS10} introduced the parameter 
\be
\xi = \frac{\eta}{\lambda}\frac{\Delta \lambda}{\Delta \eta}
\ee
where $\lambda$ is the fluid conductivity, $\eta$ is the shear viscosity, and $\Delta$ indicates the difference of the value of the property following it at the bottom and at the top of the sample. Thus, $\xi$ is a measure of the relative sizes of the contributions to NOB conditions from $\lambda$ and $\eta$. The authors state that their data for a cell with $\Gamma = 1.0$ \cite{NS03,NS10} and $\Gamma = 4.0$ \cite{NS06a} make a transition from a state of lower to one of higher \Nu\ as $\xi$ changes in their experiments from positive values of order one to negative values of -1 or less, both states having a common scaling exponent with a value slightly above 0.3. The authors assert that the transition between the two states, where roughly $\Nu \sim \Ra^{0.5}$, should not be interpreted as the ultimate state; rather, the data  correlate well with the material parameter $\xi$.

\begin{figure}
\centerline{\includegraphics[width=5in]{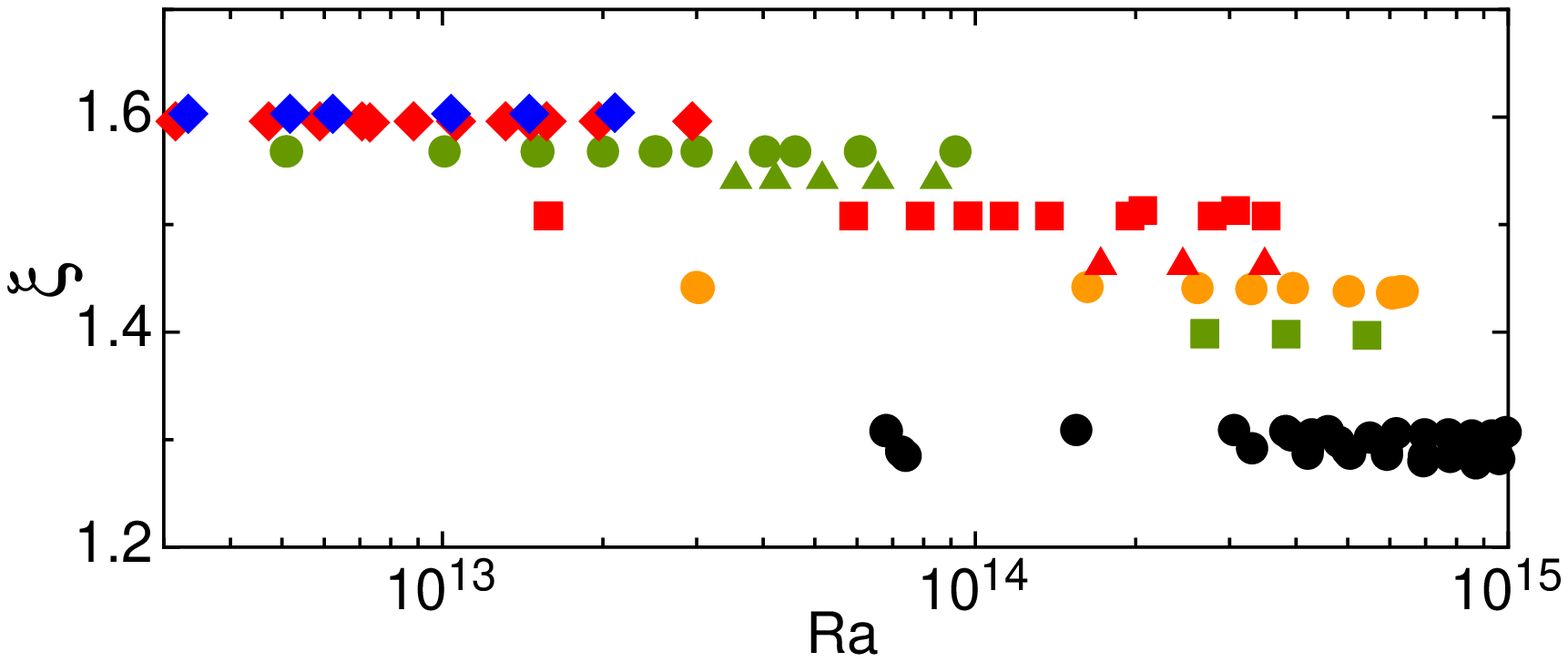}}
\caption{The parameter $\xi$ introduced by Niemela and Sreenivasan \cite{NS10} (see text). The data are for the following approximate pressures: blue diamonds: 4 bars; red diamonds: 5 bars; green circles: 8 bars; green triangles: 10 bars; red squares: 12 bars; red triangles: 14 bars; yellow circles: 15 bars; green squares: 16 bars; black circles: 18 and 19 bars.}
\label{fig:xi}
\end{figure}

The values of $\xi$ for our data are plotted in Fig.~\ref{fig:xi}. We see that at a given pressure $\xi$ is essentially independent of \Ra, and thus of $\Delta T$. The value of $\xi$ depends slightly upon the pressure, varying over the range $1.6 \agt \xi \agt 1.3$ as the pressure changes from about 4 to about 19 bars. Thus, for all of our data $\xi$ remains positive and slightly above one and the transition seen in the data of \cite{NS03,NS06a,NS10}, which were taken near the critical point of helium, is not expected to occur in our fluid-property range .

\section{Effect of horizontal temperature variations in the top and bottom plate}
\label{app:plate_T}

Maintaining the top and bottom plate at a uniform temperature, even in the presence of the large vertical heat currents up to a kW or so, is a significant experimental challenge, especially when the plates become very large as in our experiment where the diameter was well over a meter and the mass of the plates was of order 200 kg.

Our top plate was cooled by water passing through a pair of double spiral channels \cite{AFB09}. The two members of the pair were in parallel, and each pair consisted of anti-parallel ({\it i.e.} an inward and an outward flowing) spirals. The distance between the centers of adjacent channels was 2.54 cm, and the channels had a depth $d =1.26$ and a width $w = 1.27$ cm (cross sectional area of 1.60 cm$^2$). These dimensions were kept small to minimize lateral temperature variations on the small length scales of order the plate thickness. However, as will be seen below, the small channel cross section did lead to a significant flow resistance, and a somewhat larger channel cross section might have been optimal. There is a constraint on the channel depth provided by the plate thickness, and a much greater thickness would lead to excessive weight and cost of the plate. 

The plate temperatures were determined by five thermometers in each of the top plate (between the water channels) and the top of the bottom-plate composite \cite{ABFH09}. All ten thermometers were placed in small holes in the plates, with their tips within a mm or so of the copper-fluid interface. One thermometer ($T_0$) was located at the plate center. The other four $T_i$, $i = 1, ..., 4$, were equally spaced azimuthally at angular intervals of $\pi/2$ rad and were positioned radially at a distance of $0.42D$ from the center. The plate temperatures $T_b$ and $T_t$ used to calculate $\Delta T$ (and   thus \Nu\  and \Ra) were the averages of the five readings in a given plate.

\begin{figure}
\centerline{\includegraphics[width=6in]{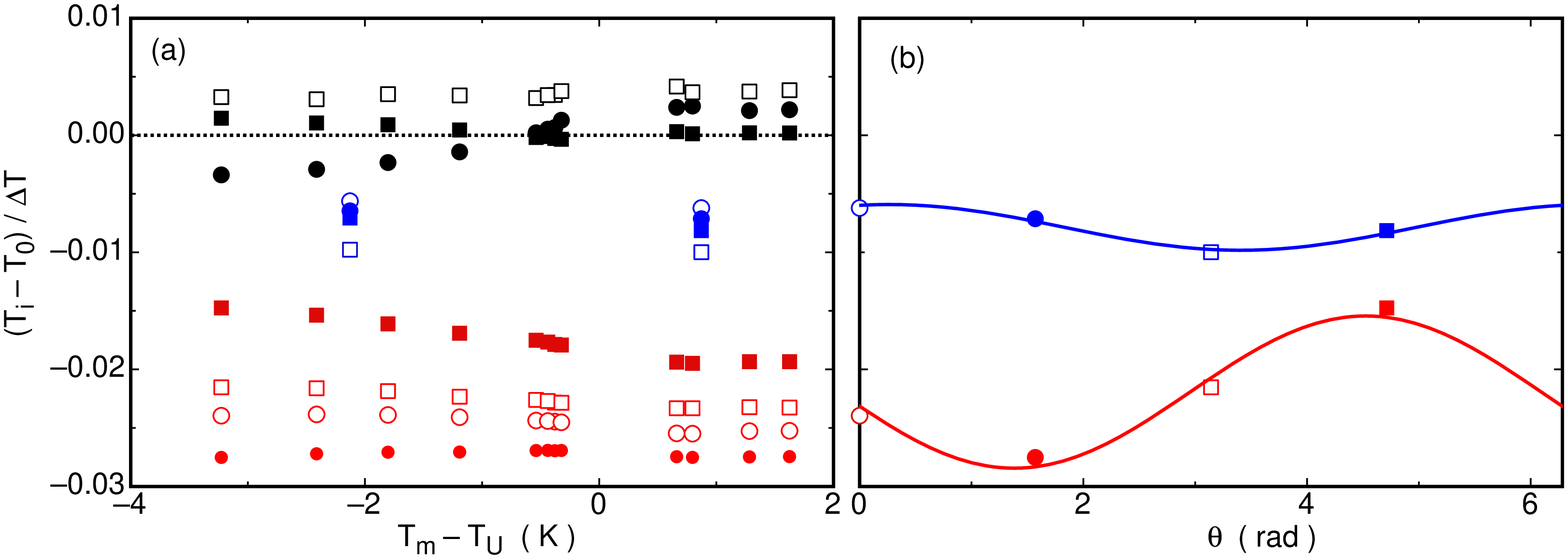}}
\caption{a.) The normalized radial temperature variation $(T_i - T_0)/\Delta T$ as a function of the temperature difference $T_m - T_U$ between the mean sample temperature and the Uboot temperature (for details, see text). Red symbols: top plate data for the weaker water circulation driven by a Nelab RTE7 circulator. Blue symbols: Data for the top plate with the stronger water circulation driven by a Wilo pump. Black symbols: Data for the top member of the bottom-plate composite. Open circles, solid circles, open squares, and solid squares are for $T_0 - T_i$ with $i = 1$, 2, 3, and 4 respectively.
b.) The azimuthal temperature variation for  two examples. The solid lines are a fit of a sine function with an adjustable phase and amplitude to the data.
}
\label{fig:TopPlateTemps}
\end{figure}

\begin{figure}
\centerline{\includegraphics[width=4in]{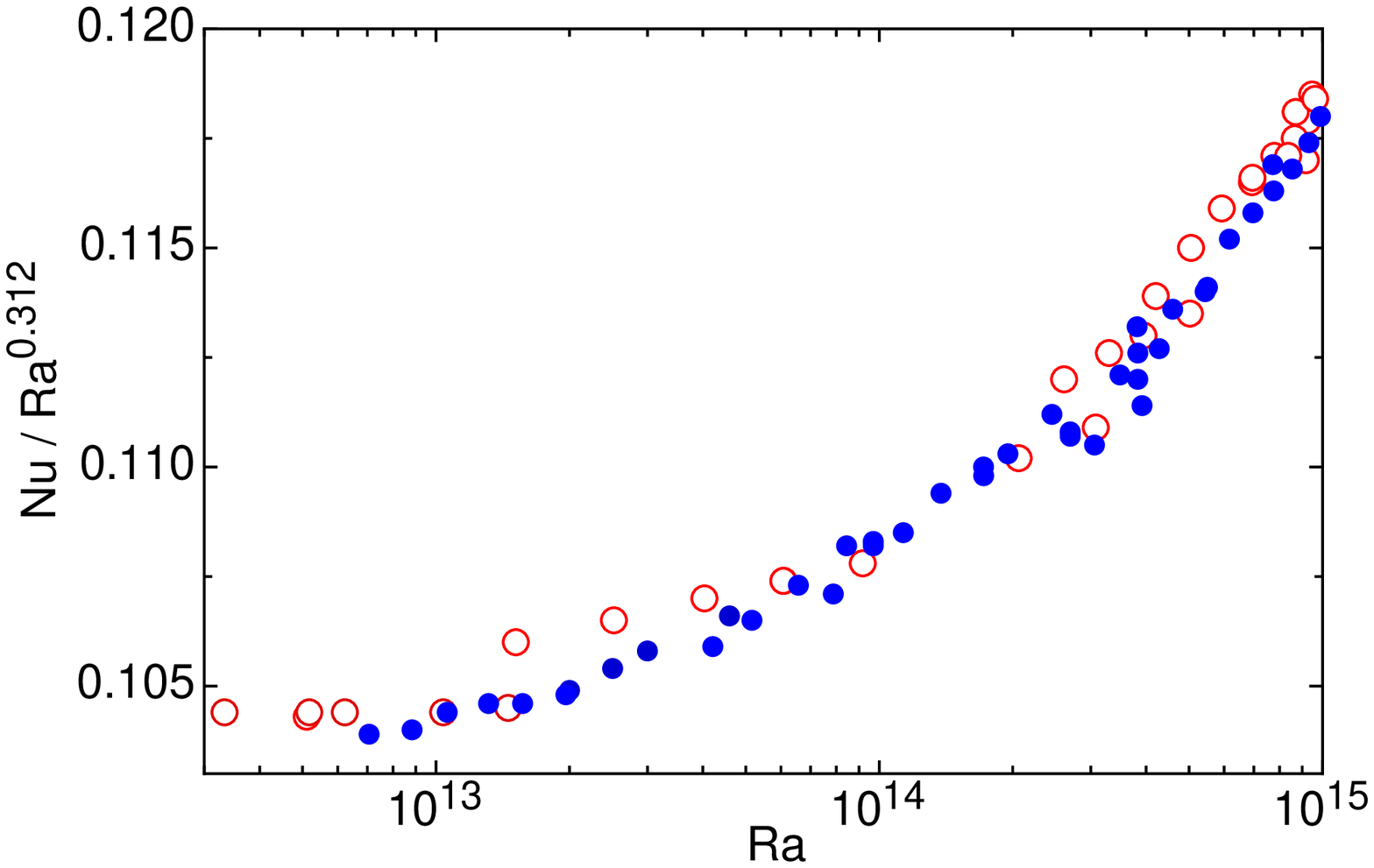}}
\caption{The reduced Nusselt number as a function of the Rayleigh number for $T_m - T_U < -2$K. Open red (solid blue) circles: data taken with low (high) water flow rate in the top-plate cooling channel.}
\label{fig:Nred_topplate}
\end{figure}

Here we report on results obtained with HPCF-IIe. Initially (run 110918 to 110410) the water flow was generated directly by a Neslab RTE7 circulator capable of generating a pump head corresponding to approximately 0.5 bars which yielded a relatively small  flow rate near 15 cm$^3$/s (0.05 m$^3$/h) in each of the two double spirals, corresponding to a mean velocity $v$ of about 10 cm/s or a Reynolds number $\Rey = v d / \nu = {\cal O}(10^3)$. Judged by the value of \Rey, the flow probably was laminar. In Fig.~\ref{fig:TopPlateTemps}a we show $(T_i - T_0)/\Delta T$, $i = 1, ..., 4$  as a function of $T_m - T_U$ as red symbols. Those data  were taken at a pressure of 18.8 bars and with $\Delta T \simeq 10$K (run 110311 to 110326) corresponding to $\Ra \simeq 8\times 10^{14}$, but similar measurements over a range of $\Delta T$ and at other pressures gave similar results. The dependence on $T_m - T_U$ is weak. The azimuthally averaged radial temperature variation $(\langle T_i \rangle_{i=1, ...,4} - T_0)/\Delta T$ is a little over two percent, which was judged larger than desirable. The azimuthal variation of  $(T_i - T_0)/\Delta T$ for $T_m - T_U = -3.2$K (run 110311) is shown in Fig.~\ref{fig:TopPlateTemps}b as red symbols. It can be fit well by a sine curve, suggesting that its origin is the large-scale circulation in the bulk fluid \cite{CCS97,BNA05,BA06a}.   

In order to increase the flow rate of the water in the top plate, we modified the top-plate water cooling circuit (starting with run 110426). We drove a primary circuit by a Wilo model MHI 205-1 pump which could generate a pressure differential across the top-plate water channel of 4.2 bars which yielded a flow rate 170 cm$^3$/s (0.6 m$^3$/h) in each double spiral, corresponding to a mean flow velocity $v \simeq 100$ cm/s or $\Rey = {\cal O}(10^4)$. This flow almost certainly was turbulent. The primary water circuit was thermally coupled by a heat exchanger to a secondary cooling circuit driven and temperature-controlled by the  Neslab RTE7 circulator. This circulator was servoed so as to maintain $T_t$ at its desired value. 
The results for $(T_i - T_0)/\Delta T$ are shown in Fig.~\ref{fig:TopPlateTemps}a as blue symbols and for two values of $T_m - T_U$. There was a significant improvement, with $(\langle T_i \rangle_{i=1, ...,4} - T_0)/\Delta T$ as small as 0.8 percent. The remaining azimuthal variation for $T_m - T_U = -2.2$ K is shown as blue symbols in Fig.~\ref{fig:TopPlateTemps}b. 

In order to judge whether one or the other of the top-plate cooling circuits is adequate to yield valid results for \Nu, we show in Fig.~\ref{fig:Nred_topplate}  results for $\Nu(\Ra)$ obtained by the first method as red open circles. Data obtained with the improved top-plate cooling method are represented by the solid blue circles. The two sets agree with each other very well and essentially within their scatter, indicating that there is no systematic dependence on the cooling method and that either method was indeed adequate for Nusselt number measurements.

The bottom plate was a composite of two copper plates with a thin Lexan layer between them. The heat current passing through this composite was generated by Joule heating with a heater wire imbedded in grooves milled into the underside of the bottom member of the composite and was uniformly distributed over the entire sample area \cite{AFB09}.  The current had to pass through the Lexan thermal barrier before passing through the top copper member and then entering the fluid. Results for $(T_i - T_0)/\Delta T$ obtained for runs 110918 to 110410 are shown as black symbols in Fig.~\ref{fig:TopPlateTemps}a. Their average values are close to zero, showing that the radial temperature variation in the bottom plate was remarkably small and indeed negligible.

\section{Evidence for a closed sample}
\label{app:closed}

\begin{figure}
\centerline{\includegraphics[width=5in]{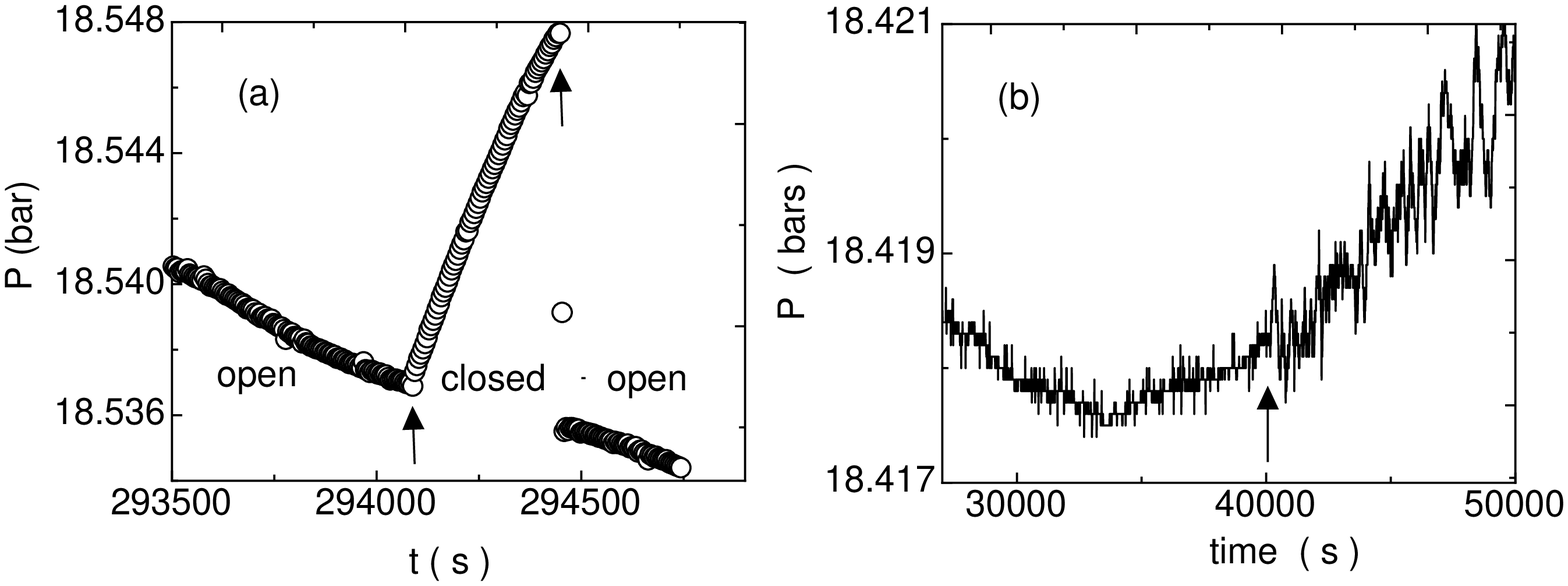}}
\caption{a): The sample pressure as a function of time. The data were taken on Sep. 14 2010, during a time interval when the Uboot temperature and pressure had not yet fully equilibrated. The sharp increase at the time indicated by the first arrow occurred when the valve was closed, and the sudden decrease at the second arrow was caused by opening the valve again. b): The sample pressure as a function of time since the beginning of run 100918. Initially, the valve was open and the sample pressure could equilibrate with the Uboot pressure. When the valve was closed at $t = 40200$ s (the location of the arrow), the pressure could no longer equilibrate with the Uboot pressure and the fluctuations due to the turbulent convection became noticeable. This run is for $T_m = 20.69^\circ$C, 18.4 bars,  and $\Delta T = 11.37$K, corresponding to  $\Ra = 9.25\times 10^{14}$. }
\label{fig:closedcell}
\end{figure}

HPCF-IIe was supposed to be completely sealed, except for a 2.5 cm diameter pipe entering it at half-height through the side wall. This pipe terminated outside the sample in a remotely operable valve which was to be open during an equilibration period to allow pressure equilibration between the Uboot and the sample, but closed when the system had reached a steady state after a new set point of the top- and bottom-plate temperatures had been established. In Fig.~\ref{fig:closedcell}a we show the pressure inside the sample during this equilibration period. With the valve open, the pressure of the Uboot and sample still drifted at a significant rate. In this case, however, the valve was closed prematurely. One can see that the sample pressure rose. When the valve was opened again, the pressure returned back to the Uboot pressure which all along had been drifting downward, albeit at a lesser rate. We regard this measurement as evidence that sealing the sample was indeed successful.

Figure~\ref{fig:closedcell}b shows the pressure with higher resolution after a steady state had been reached. With the valve open, the noise level of the readings, reflecting the instrumental noise,  is quite small but noticeable. After closing the valve, the noise level increases because the genuine noise in the pressure due to the turbulent convection in the sample becomes noticeable because it can no longer be smoothed out by exchange of fluid with the Uboot. This phenomenon also indicates that sealing the sample was successful.

\section{Effect of side-shield mismatch with the sample temperature}
\label{app:sideshield}

 \begin{table}[h]
\caption{Effect of a side-shield temperature-deviation from the mean temperature for run 120415. $\langle T_{ss} \rangle$ is the set-point of the shield temperature. $T_{SS}(iL/4)$, $i = 1, 2, 3$, are the measured shield temperatures at the three vertical positions $iL/4$ as measured from the sample bottom.}
\vskip 0.1in
\begin{center}
\begin{tabular}{ccc}
Quantity & $\langle T_{SS}\rangle = T_m$ & $\langle T_{SS}\rangle = T_m + 1 {\rm K}$ \\  
\hline
$T_m$   &   21.587&   21.599 \\
$T_c$   &  21.398 &  21.441  \\
$\Delta T$   &10.163   &   10.191 \\
$T_{SS}(L/4)$   & 21.629  & 22.621   \\
$T_{SS}(2L/4)$   &  21.602 & 22.602   \\
$T_{SS}(3L/4)$   & 21.550  & 22.571   \\
$Q$ [Watts]   & 320.6  & 319.6   \\
$\Ra$   & $ 6.70\times 10^{14}$ & $ 6.76\times 10^{14}$  \\
$\Nu$   &  4779 &  4747  \\

\end{tabular}
\end{center}
\label{tab:SSoffset}
\end{table}

\begin{figure}
\centerline{\includegraphics[width=5in]{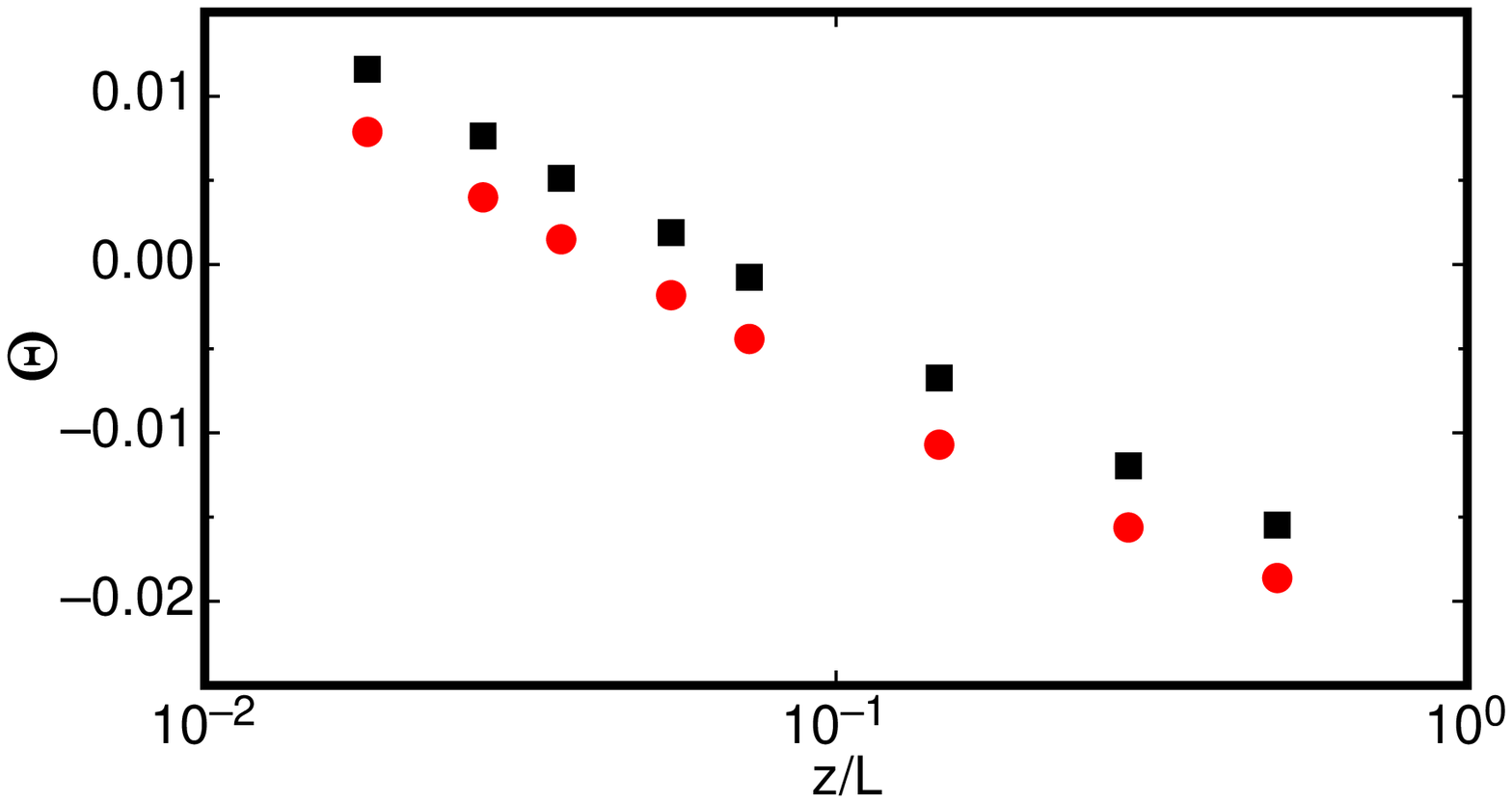}}
\caption{The fluid temperature at a radial distance of one cm from the side wall as a function of the height $z/L$ above the bottom of the sample. The red circles (black squares) are for $\langle T_{SS}\rangle = T_m$ ($\langle T_{SS}\rangle = T_m + 1$ K).}
\label{fig:T_v}
\end{figure}

As shown in Fig.~\ref{fig:samplecell}, the side wall of the sample is surrounded by a side shield (``SS" in the figure) which is intended to prevent heat loss or input through the side wall. The temperature of this shield was controlled at the mean temperature $T_m$. Since the sample center temperature $T_c$ will differ slightly from $T_m$ due to non-Boussinesq effects, it is important to ask whether heat input to the side wall due to the temperature difference $T_m - T_c$ is significant. Further, in view of the large size and mass of the shield, it is appropriate to ask how successful its temperature control was at a specified setpoint. Thus, a run was conducted  in which the shield temperature was deliberately servoed at a  displacement from $T_m$ which is much larger than $T_m-T_c$ or any deviations from perfect temperature regulation, namely at $T_m + 1.0$ K. Various measured quantities with both setpoints are shown in Table~\ref{tab:SSoffset}. This example is for a nominal $\Delta T = 10.1$ K.

First we call attention to the shield temperatures $T_{SS}$ at the three vertical positions $L/4,~L/2,~3L/4$ (the sample height $L$ was 2.2 m) measured from the bottom plate. They span a vertical distance of over one meter. One sees that the temperature gradient in the shield was typically about 0.05 K/m, which suggests that the shield temperature was uniform throughout within about 0.1 K.

A small shift of $T_m$ and $\Delta T$ can be seen to have occurred due to the shift of the shield temperature by one K. This could happen because the bottom plate is a composite with a lower (BPb) and an upper (BPt) member (see Fig.~\ref{fig:samplecell}), with the temperature of BPb controlled. Thus the temperature of the BPt could adjust itself to prevailing heat currents and influence $T_m$ as well as $\Delta T$. The effect of a one K change in the shield temperature caused a change of 28 mK, or 0.3\%, of $\Delta T$ and a shift of $T_m$ by 12 mK. The effect of imperfect shield regulation on these quantities can be regarded as negligible.

The Rayleigh number was shifted by about  0.9\%. This is due in part to the change of $\Delta T$ and in part due to a change of the fluid properties associated with the shift of $T_m$. Since actual temperature offsets are much less than one K, this effect is not serious. Similarly \Nu\ is affected, by about 0.67\%. This shift is due both to the change of $\Delta T$ and to a small change of the heat current needed to maintain the temperature of the bottom member of the bottom-plate composite (BPb) at the specified set-point. Again this is not a serious shift.  

For completeness we also discuss the influence of the shield temperature on the sample temperature near the side wall. Although not directly relevant to the present paper, it is important for other related investigations of temperature profiles in the bulk of the sample \cite{ABGHLSV12}. For that purpose eight thermometers were mounted in the sample at a radial position one cm from the side wall and at eight vertical positions $z_j$. The eight temperatures $\Theta(z_j) \equiv [\langle T(z_j )\rangle - T_m] / (T_b - T_t)$  (we denote the time average by $\langle ... \rangle$) are shown in Fig~\ref{fig:T_v} as a function of $z/L$. One sees that there is no change in the shape of the vertical temperature profile, but there is a small shift as had been indicated already by the shift of $T_c$ shown in the table. 

Finally we note that there are several other thermal shields in the system (see Sec.~\ref{sec:apparatus}), but we believe that their influence on the system performance is smaller than that of the side shield.
\vfill\eject

\section{Data tables}
\label{app:data}
  
 \begin{table}[h]
\caption{SF$_6$, HPCF-IIe, level sample (tilt angle $\beta < 0.0001$ rad).}
\vskip 0.1in
\begin{center}
\begin{tabular}{cccccccc}

Run No.&	$P$ (bars)&	$T_m (^\circ{\mathrm C})$&	$T_U (^\circ{\mathrm C})$&	$\Delta T$ (K)&	\Ra&	\Pra&	\Nu\\
\hline
100918&		18.419&	20.691&	23.777&	11.370&	9.25IIe+14&	0.861&	5507.45\\
100924&		18.466&	20.788&	24.286&	11.564&	9.477e+14&	0.862&	5579.63\\
100925&		18.502&	21.844&	24.446&	11.679&	9.167e+14&	0.862&	5447.81\\
101004&		8.011&	20.896&	24.343&	11.786&	6.086e+13&	0.799&	2145.55\\
101025&		8.000&	20.973&	23.993&	2.947&	1.514e+13&	0.799&	1371.90\\
101027&		8.003&	20.950&	24.101&	4.896&	2.519e+13&	0.799&	1616.40\\
101029&		7.998&	20.925&	23.947&	7.846&	4.03IIe+13&	0.799&	1879.88\\
101031&		8.005&	20.895&	24.160&	11.786&	6.075e+13&	0.799&	2146.16\\
101102&		8.016&	20.870&	24.549&	17.732&	9.179e+13&	0.799&	2449.66\\
101114&		11.751&	20.977&	24.611&	13.447&	2.063e+14&	0.820&	3224.14\\
101116&		11.762&	21.021&	24.849&	20.030&	3.079e+14&	0.821&	3675.61\\
101108&		14.891&	20.920&	25.015&	17.826&	6.060e+14&	0.842&	4659.07\\
101109&		14.894&	21.247&	25.008&	18.479&	6.21IIe+14&	0.842&	4694.89\\
101120&		14.894&	21.382&	24.994&	18.749&	6.27IIe+14&	0.842&	4710.51\\
101122&		14.843&	21.498&	24.246&	11.988&	3.945e+14&	0.841&	4045.17\\
101124&		14.810&	21.272&	23.786&	7.938&	2.61IIe+14&	0.841&	3526.58\\
101126&		14.792&	21.282&	23.526&	4.960&	1.624e+14&	0.841&	2995.86\\
101130&		14.802&	20.970&	23.740&	9.934&	3.298e+14&	0.841&	3814.08\\
101202&		14.839&	21.011&	24.271&	15.011&	5.021e+14&	0.841&	4382.99\\
101204&		14.898&	21.517&	25.061&	19.021&	6.339e+14&	0.842&	4724.05\\
101207&		14.873&	20.978&	24.757&	17.942&	6.059e+14&	0.842&	4661.48\\
101214&		18.739&	20.933&	23.730&	11.523&	1.004e+15&	0.863&	6154.13\\
101215&		18.769&	20.971&	23.918&	9.934&	8.699e+14&	0.863&	5410.40\\
101217&		18.755&	20.973&	23.781&	7.945&	6.93IIe+14&	0.863&	4975.08\\
101218&		18.814&	21.644&	24.306&	12.274&	1.051e+15&	0.863&	5871.83\\
101219&		18.817&	21.612&	24.295&	11.214&	9.620e+14&	0.863&	5599.07\\
101221&		18.812&	21.536&	24.246&	10.065&	8.657e+14&	0.863&	5375.00\\
101223&		18.785&	21.566&	23.998&	9.125&	7.785e+14&	0.863&	5183.77\\
101225&		18.781&	21.583&	23.960&	8.160&	6.949e+14&	0.863&	4981.14\\
101227&		18.751&	21.490&	23.682&	6.975&	5.924e+14&	0.863&	4710.15\\
101229&		18.709&	21.432&	23.289&	0.861&	7.258e+13&	0.863&	1754.60\\
101231&		18.740&	21.483&	23.570&	5.961&	5.051e+14&	0.863&	4448.59\\
110102&		18.735&	21.487&	23.522&	4.971&	4.205e+14&	0.863&	4160.20\\
\end{tabular}
\end{center}
\label{tab:level}
\end{table}

\vfill
\eject

\begin{table}[h]
\caption{SF$_6$, HPCF-IIe, First tilted sample, tilt angle $\beta = 0.003$ rad.}
\vskip 0.1in
\begin{center}
\begin{tabular}{cccccccc}

Run No.&	$P$ (bars)&	$T_m (^\circ{\mathrm C})$&	$T_U (^\circ{\mathrm C})$&	$\Delta T$ (K)&	\Ra&	\Pra&	\Nu\\
\hline

110115&		4.181&	20.950&	23.734&	4.896&	5.176e+12&	0.787&	966.79\\
110117&		4.187&	20.931&	24.142&	19.859&	2.107e+13&	0.787&	1497.77\\
110129&		4.180&	21.083&	23.621&	3.164&	3.334e+12&	0.787&	843.02\\
110131&		4.187&	20.930&	24.159&	19.858&	2.107e+13&	0.787&	1498.53\\
110202&		4.184&	20.855&	23.987&	13.708&	1.455e+13&	0.787&	1335.73\\
110204&		4.184&	20.895&	23.947&	9.791&	1.038e+13&	0.787&	1201.21\\
110206&		4.185&	20.938&	24.046&	5.875&	6.228e+12&	0.787&	1024.77\\
110221&		18.565&	21.040&	24.198&	12.067&	1.001e+15&	0.862&	5707.16\\
110227&		18.574&	21.034&	24.286&	12.057&	1.00IIe+15&	0.862&	5716.32\\
110311&		18.574&	21.034&	24.260&	10.059&	8.364e+14&	0.862&	5302.69\\
110417&		18.603&	21.036&	24.563&	7.070&	5.921e+14&	0.862&	4694.05\\
110418&		18.572&	21.035&	24.259&	5.068&	4.211e+14&	0.862&	4133.38\\
110419&		18.554&	20.996&	24.093&	3.990&	3.307e+14&	0.862&	3758.72\\
110420&		18.573&	20.994&	24.283&	5.990&	4.990e+14&	0.862&	4412.69\\
110421&		18.620&	22.003&	24.600&	6.004&	4.813e+14&	0.863&	4374.31\\
110524&		11.837&	24.996&	28.197&	13.985&	1.956e+14&	0.820&	3157.24\\
110525&		11.827&	23.264&	28.250&	15.655&	2.294e+14&	0.821&	3336.36\\
110526&		11.828&	23.974&	28.184&	15.941&	2.289e+14&	0.821&	3331.77\\
110526&		11.820&	23.959&	28.025&	11.912&	1.708e+14&	0.820&	3023.89\\
110527&		11.809&	23.961&	27.803&	7.918&	1.13IIe+14&	0.820&	2633.26\\

\end{tabular}
\end{center}
\label{tab:tilt1}
\end{table}

\vfill\eject

 \begin{table}[h]
\caption{SF$_6$, HPCF-IIe, Second tilted sample, tilt angle $\beta = 0.014$ rad.}
\vskip 0.1in
\begin{center}
\begin{tabular}{cccccccc}

Run No.&	$P$ (bars)&	$T_m (^\circ{\mathrm C})$&	$T_U (^\circ{\mathrm C})$&	$\Delta T$ (K)&	\Ra&	\Pra&	\Nu\\
\hline

110625&		4.971&	24.017&	27.841&	1.033&	1.533e+12&	0.789&	661.28\\
110629&		6.953&	23.855&	28.251&	19.704&	6.607e+13&	0.795&	2178.61\\
110630&		6.946&	23.908&	27.964&	9.814&	3.279e+13&	0.795&	1744.85\\
110701&		6.943&	23.953&	27.841&	4.904&	1.635e+13&	0.795&	1395.79\\
110716&		18.237&	20.992&	23.753&	4.981&	3.817e+14&	0.861&	4011.57\\
110717&		18.275&	20.998&	24.135&	9.990&	7.725e+14&	0.861&	5162.00\\
110718&		18.282&	20.997&	24.244&	12.980&	1.006e+15&	0.861&	5707.97\\
110718&		18.303&	21.524&	24.363&	13.033&	9.894e+14&	0.861&	5629.48\\
110720&		18.383&	21.531&	25.210&	13.647&	1.056e+15&	0.862&	5790.57\\
110722&		18.379&	21.530&	25.156&	13.046&	1.009e+15&	0.862&	5673.03\\
110723&		18.379&	21.531&	25.135&	12.050&	9.316e+14&	0.862&	5496.51\\
110724&		18.384&	21.523&	25.183&	11.037&	8.548e+14&	0.862&	5323.81\\
110725&		18.381&	21.515&	25.144&	10.023&	7.759e+14&	0.862&	5144.72\\
110726&		18.370&	21.512&	25.037&	9.019&	6.965e+14&	0.861&	4950.18\\
110727&		18.357&	21.507&	24.906&	8.007&	6.166e+14&	0.861&	4741.89\\
110728&		18.332&	21.500&	24.654&	6.000&	4.59IIe+14&	0.861&	4264.08\\
110731&		18.340&	21.498&	24.760&	4.993&	3.830e+14&	0.861&	3973.54\\
110801&		18.323&	21.004&	24.672&	5.004&	3.916e+14&	0.861&	3980.43\\
110802&		18.341&	21.001&	24.848&	6.997&	5.501e+14&	0.861&	4531.63\\
110803&		18.339&	21.261&	24.791&	5.520&	4.281e+14&	0.861&	4140.67\\
110803&		16.166&	21.475&	24.995&	5.945&	2.696e+14&	0.850&	3521.90\\
110804&		16.168&	21.475&	25.014&	5.946&	2.697e+14&	0.850&	3521.17\\
110805&		16.173&	21.467&	25.078&	8.431&	3.831e+14&	0.850&	3996.65\\
110806&		16.182&	21.473&	25.197&	11.937&	5.435e+14&	0.850&	4512.01\\
110807&		14.105&	21.649&	24.915&	6.295&	1.720e+14&	0.836&	3035.56\\
110808&		14.105&	21.650&	24.909&	6.291&	1.719e+14&	0.836&	3039.94\\
110809&		14.113&	21.742&	25.037&	8.982&	2.45IIe+14&	0.836&	3431.72\\
110810&		14.136&	21.899&	25.354&	12.790&	3.49IIe+14&	0.836&	3863.76\\
110811&		12.061&	21.453&	25.045&	5.906&	9.694e+13&	0.822&	2503.66\\
110812&		12.060&	21.454&	25.034&	5.908&	9.695e+13&	0.822&	2500.66\\
110813&		12.065&	21.442&	25.135&	8.380&	1.378e+14&	0.822&	2822.21\\
110814&		12.067&	21.434&	25.177&	11.861&	1.951e+14&	0.822&	3171.61\\
110815&		12.078&	21.445&	25.383&	16.879&	2.784e+14&	0.823&	3564.35\\
\end{tabular}
\end{center}
\end{table}

\vfill\eject

 \begin{table}[h]
\begin{center}
\begin{tabular}{cccccccc}
110816&		12.101&	21.673&	25.806&	21.336&	3.515e+14&	0.823&	3839.08\\
110817&		12.077&	21.437&	25.356&	6.872&	1.133e+14&	0.823&	2631.68\\
110824&		10.068&	21.454&	25.455&	5.404&	5.161e+13&	0.810&	2021.40\\
110625&		10.068&	21.440&	25.457&	6.877&	6.570e+13&	0.810&	2196.23\\
110626&		10.063&	21.425&	25.319&	8.846&	8.440e+13&	0.810&	2395.76\\
110827&		8.013&	21.510&	25.143&	9.017&	4.593e+13&	0.799&	1950.67\\
110828&		8.010&	21.444&	25.047&	5.885&	2.999e+13&	0.799&	1695.53\\
110829&		8.013&	21.454&	25.137&	4.908&	2.503e+13&	0.799&	1597.10\\
110911&		5.025&	21.422&	25.492&	6.604&	1.060e+13&	0.789&	1209.78\\
110912&		5.023&	21.390&	25.369&	9.779&	1.568e+13&	0.789&	1369.93\\
110913&		5.023&	21.372&	25.387&	12.240&	1.964e+13&	0.789&	1471.19\\
110914&		5.020&	21.407&	25.206&	8.205&	1.314e+13&	0.789&	1295.76\\

\end{tabular}
\end{center}
\label{tab:tilt2}
\end{table}

\vfill\eject



\end{document}